\documentclass[12pt]{article}

\usepackage{scicite}

\usepackage{times}
\usepackage{graphicx}
\usepackage{xcolor}
\usepackage{comment}

\newenvironment{sciabstract}{%
\begin{quote} \bf}
{\end{quote}}

\title{Microbes in porous environments: \\ From active interactions to emergent feedback}

\author
{Chenyu Jin$^{1}$ and Anupam Sengupta$^{1,2\ast}$ \\
\\
\normalsize{$^{1}$ Physics of Living Matter Group, Department of Physics and Materials Science,}\\
\normalsize{University of Luxembourg, 162 A, Avenue de la Faïencerie, L-1511 Luxembourg City, Luxembourg}\\
\\
\normalsize{$^{2}$ Institute for Advanced Studies, University of Luxembourg,}\\
\normalsize{2 Avenue de l’Université, L-4365, Esch-sur-Alzette, Luxembourg}\\
\\
\normalsize{$^\ast$To whom correspondence should be addressed; E-mail:  anupam.sengupta@uni.lu}\\
}

\date{}

\begin{document} 


\baselineskip24pt


\maketitle


\begin{sciabstract}
  Microbes thrive in diverse porous environments--from soil and riverbeds to human lungs and cancer tissues--spanning multiple scales and conditions. Short- to long-term fluctuations in local factors induce spatio-temporal heterogeneities, often leading to physiologically stressful settings. How microbes respond and adapt to such biophysical constraints is an active field of research where considerable insight has been gained over the last decades. With a focus on bacteria, here we review recent advances in self-organization and dispersal in inorganic and organic porous settings, highlighting the role of active interactions and feedback that mediate microbial survival and fitness. We conclude by discussing open questions and opportunities for using integrative cross-disciplinary approaches to advance our understanding of the biophysical strategies that microbes employ--at both species and community scales--to make porous settings habitable. Active and responsive behaviour is key to microbial survival in porous environments, with far-reaching ramifications for mitigating anthropogenic impacts, innovating subsurface storage solutions, and predicting future ecological scenarios under current climatic changes.  
  
\end{sciabstract}


\section*{Introduction}

Microorganisms including bacteria, archaea, fungi and diverse protists, have successfully inhabited the biosphere, pervading the several kilometers into the atmosphere, the oceans, the seabed, and the depths of subsurface rocks. With the total number of prokaryotes (bacteria and archaea) on Earth estimated to be about $1.2 \times 10^{30}$ cells \cite{flemming2019bacteria}, they account for nearly $15\%$ of the net biomass on Earth, despite being orders of magnitude smaller size than those of the eukaryotic cells \cite{bar2018biomass, philippot2023}. Four of the ``big five'' habitats where majority of microbes are found: soil, oceans, deep continental subsurface, deep oceanic subsurface and the upper oceanic sediments, are of a porous nature and host around $90\%$ of all microorganisms \cite{ebrahimi2014, dang2016, flemming2019bacteria}. In addition, microbes often inhabit living and decaying plant and animal bodies, exploiting symbiotic or pathogenic relationships. Such organic substrates, including mucus linings \cite{kirch2012, bansil2013, bansil2018, wu2023mucin}, food materials \cite{aminifar2010, ranjbaran2021, dadmohammadi2022}, and flocculated suspended sediments (flocs) including marine snow \cite{kiorboe2002, nguyen2017, lawrence2023, borer2023}, offer dynamic porous environments where microstructure and local viscoelasticity mediate microbial dispersal, locomotion and colonization over time. 

Despite their tiny size, microbes are evolutionarily well equipped to disperse widely, relying on a range of passive and active mechanisms. Taking \textit{Bacillus subtilis} as an example, this common soil bacterium is only a few microns in size, but manages to colonize soil across all continents \cite{philippot2023} 
and in underground water columns several kilometers deep into the soil \cite{kondakova2013study}. Microbial communities inhabiting porous subsurface environments have received particular attention in recent years due to their implications in hydrogen (H$_{2}$) storage. H$_{2}$, an important potential replacement of fossil fuels, has necessitated the search for new storage facilities in the form of depleted gas fields and aquifers \cite{heinemann2021, muhammed2022}. However, our understanding of the behaviour and impacts of stored H$_{2}$ within such porous settings, particularly in the context of a diverse microbial ecosystem thriving in these environments, is at a nascent stage \cite{thaysen2021, liu2023}. Due to the anoxic conditions, subsurface microbial communities may utilize H$_{2}$ as an electron donor, producing various gaseous by-products including hydrogen sulfide and methane. Biological activity and microbial growth over time lead to the formation of biofilms, with subsequent clogging of the subsurface pores and a significant increase in pressures required for gas injection \cite{eddaoui2021}. In addition, gaseous metabolites generated by microbes can alter the pore-scale wettability, while interaction with water molecules can trigger the formation of acidic compounds, leading to behavioural and physiological impacts on the microbial communities \cite{liu2019,santha2019,pan2022}.

\begin{figure*}
  \centering
  \includegraphics[width=\textwidth]{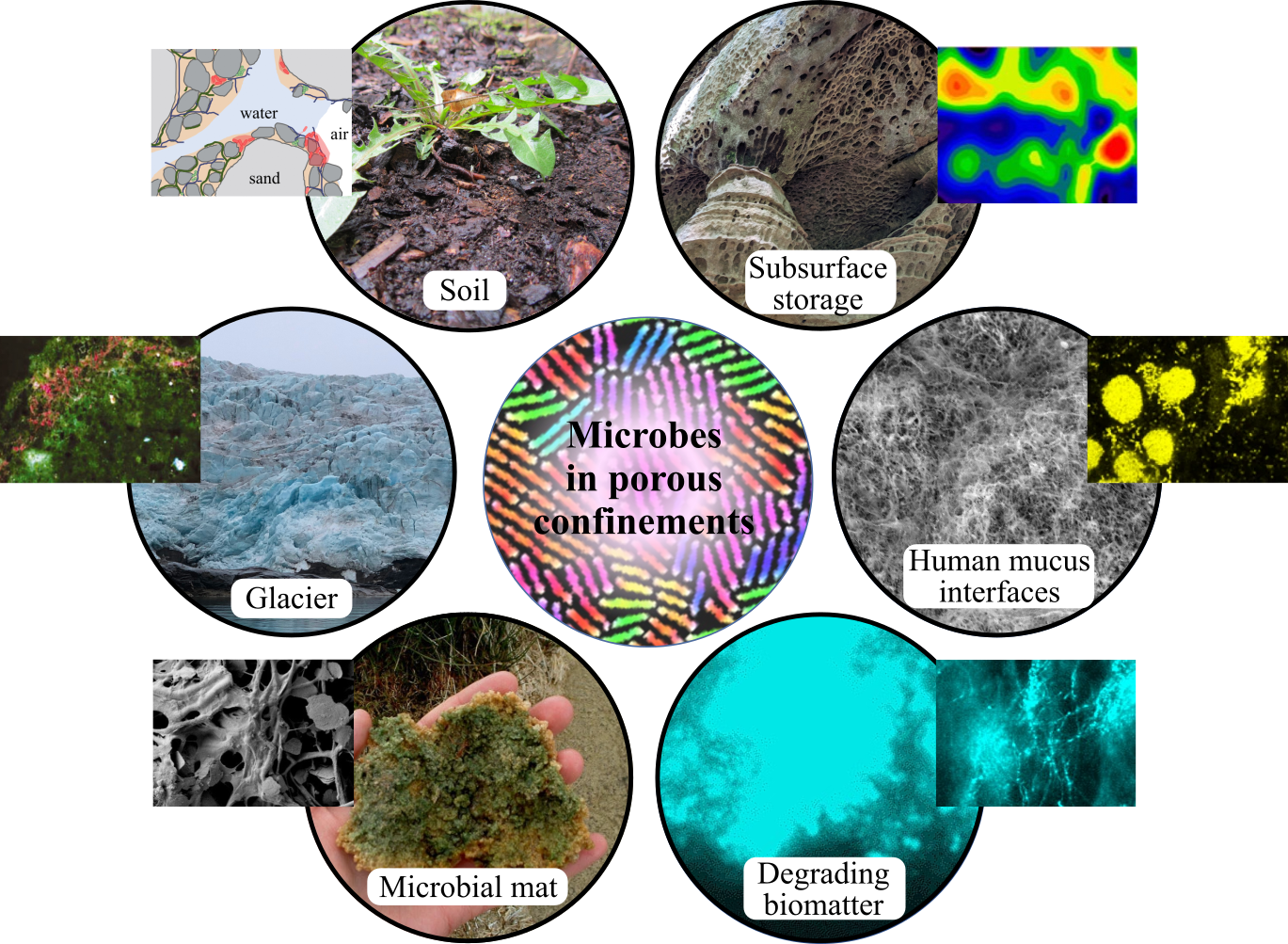}
  \caption{Microbes inhabiting porous environments. Clockwise from top: Soil is a packed, organo-mineral complex. Side panel shows the microscopic inner structure of soil. 
  Subsurface main panel shows honeycomb patterned sandstone in Mullerthal region, Luxembourg. Subsurface inner structure showing contour plots of hydrogen gas saturation in the pore network during bacterial growth, adapted from \cite{liu2023pore}, available under the terms of CC BY 4.0.
  Human mucus found in the gut, lungs and urinary tract offer habitats to diverse bacterial species. Side panel shows the fluorescence in situ hybridization (FISH) image on biofilms formed by native salivary communities over 48 hours, adapted from \cite{wu2023mucin}, CC BY 4.0.
  Degrading planktonic matter represents a hierarchical fibrous structure (shown in cyan), acting as a chemo-attractant for motile bacterial species.
    A benthic microbial mat found in the wetland. The side panel shows the SEM image of the internal structure of the microbial mat, mineral particles associated with filamentous cyanobacteria. Panel adapted from \cite{power2007biologically}, CC BY 2.0.
    Glacier main panel is taken at Billefjorden, Norway by Xinran O. Zhao. Side panel shows the inner structure of a cryoconite aggregates, red indicates cynobacteria under UV), adapted with permission from Reference \cite{hodson2010structure}, copyright 2010 Cambridge University Press.
  }
  \label{fig:porous-exp}  
\end{figure*}

Microbes disperse through porous confinements by both passive pathways \cite{conrad2018confined, rhodeland2020} and active biophysical mechanisms \cite{tecon2017biophysical, scheidweiler2020, balseiro-romero2022}. Passive pathways include convective transport due to fluid flow, attachment and subsequent transport by soil particles or hitchhiking larger organisms inhabiting pore spaces, such as worms, nematodes, and insects. Active dispersal mechanisms, typically under hydrated conditions, may involve surface-mediated movement by swarming, twitching, gliding or sliding, often mediated by physico-chemical gradients that exist or develop locally. Due to the microscale body length of bacteria, their active dispersal is spatially limited \cite{wisnoski2023scaling, wu2023}. To date, most studies of microbial ecology in porous environments, particularly in the context of biogeography, have primarily considered passive dispersal. However, several recent papers have shown how active dispersal, aided by the dynamic transformation of microbial life forms between sessile and planktonic states could play a critical role in microbial colonization, interactions and community-scale resilience in diverse porous confinements \cite{mcdougald2012, smith2018, coyte2017microbial, taubert2018, abiriga2022, soares2023}. 

In this review, we present key advances in our understanding of active microbial strategies within diverse porous environments (covering last 15 years), with an overarching aim to identify universal local properties and mechanisms which microbes put to use across different microstructural and physico-chemical constraints. 
This review is organised as follows:
In Section II, we provide a brief overview of different microbe-inhabited porous environments, and the associated constraints they offer for microbial dispersal and locomotion.
In Section III, we present recent literature on active strategies of bacterial dispersal, with relevance to other microbial families including archaea, algae, fungi. 
Finally, in Section IV, we offer a perspective highlighting some of the key open questions and challenges in the field of microbial organization in porous environments.

\section*{Porous microbial habitats}


\begin{table*}
  \caption{\label{tab:porous} Physical overview of a selection of microbe-inhabited porous environments.}
  
    \begin{tabular}{ccccc}
      \\
      Porous environment & Structure & Structural stability & Water content \\ \hline
      Soil & loose/tight packing & labile 
      & unsaturated/saturated & \\ 
      Subsurface rocks & tight packing & stable & unsaturated/saturated & \\ 
      Cryoconite & loose/tight packing & labile & saturated & \\ 
      Microbial mats & fibrous/hierarchical & labile & saturated & \\ 
      Degrading biomatter & fibrous/hierarchical & labile & saturated & \\
      Healthy tissues & fibrous/hierarchical & labile & saturated & \\ 
      Cancerous tissues & fibrous/hierarchical & stiff & unsaturated/saturated & \\ 
      Hydrogels (synthetic) & fibrous/gel & labile & saturated & \\ \hline
    \end{tabular}
\end{table*}

\paragraph*{Soil.}  Soil contains about one quarter of total prokaryotic cells concentrated in a small volume \cite{flemming2019bacteria}, hosting approximately $3 \times 10^{29}$ prokaryotic cells. 
Around $50\%$ by volume of topsoil is solid, with the remaining half typically filled with water and air in a dynamic equilibrium. The water (and the relative air) content varies with prevailing environmental and weather conditions \cite{konig2020physical, vos2013micro}. 
While the production of extracellular polymeric substances often leads to bioclogging of the porous networks, they also enable water retention keeping soils alive and fertile \cite{philippot2023, volk2016, rabbi2020microbial}. The intrinsic structural heterogeneity, offered by the packed, organo-mineral complex and hierarchical porosity, provides suitable niches for different microbial species \cite{philippot2023}. Sand, silt and clay--the primary constituents--aggregate to form small clusters ($<20\,\mu\mathrm{m}$), which then bound together to form larger microaggregates ($20$--$250\,\mu\mathrm{m}$), and finally into the loosely packed macroaggregates ($>250\,\mu\mathrm{m}$) \cite{totsche2018microaggregates}. While the microaggregates are mechanically robust, held together mainly by the van der Waals and electrostatic forces, the macroaggregates are relatively labile being loosely held together by polysaccharides, root and fungal hyphae. Overall, the soil body--an aggregation of macroaggregates--is an unstable structure, susceptible to various physico-chemical changes over daily to seasonal timescales, resulting in a heterogeneous and fluctuating microbial distribution \cite{tecon2017biophysical}.

\paragraph*{Subsurface rocks.} The subsurface refers to area of the Earth's crust and deep sea sediments that are estimated to be several kilometers thick. 
Despite the extreme environmental conditions, characterized by lack of light, lack of oxygen, and high pressures and temperatures, they harbor active microbial communities \cite{escudero2023}. 
Although low in density and metabolism, microorganisms in the continental and oceanic subsurface account for more than half of total microbial biomass \cite{flemming2019bacteria}, contributing to biogeochemical processes including diagenesis, weathering, precipitation, and redox reactions of minerals. 
In particular, the presence of water in such extreme environments (made possible by fractured and highly porous rocks) promote colonisation of microorganisms. A recent study has reported the existence of multi-species biofilm communities at depths between $139\,\mathrm{m}$ and $519\,\mathrm{m}$, with potentially altered metabolic pathways compared to their counterparts thriving under normal environmental conditions \cite{escudero2018active}. Subsurface rocks, together with aquifers, recently attracted attention for their potential use as H$_{2}$ reservoirs \cite{heinemann2021,tarkowski2022towards}. However, systematic studies are needed to understand the interactions and feedback between H$_{2}$ and the subsurface microbial community. As an electron donor, H$_{2}$ is essential for the survival of anoxic species which leverage acetogenesis, sulfate reduction and methanogenesis for survival \cite{harris2007situ}. H$_{2}$-mediated metabolic processes may alter the pore-scale properties \cite{liu2023}, thus long-term monitoring and testing will be critical in assessing the suitability of subsurface rocks as reservoirs \cite{heinemann2021,thaysen2021}.



\begin{figure*}
  \centering
  \includegraphics[width= 0.6\columnwidth]{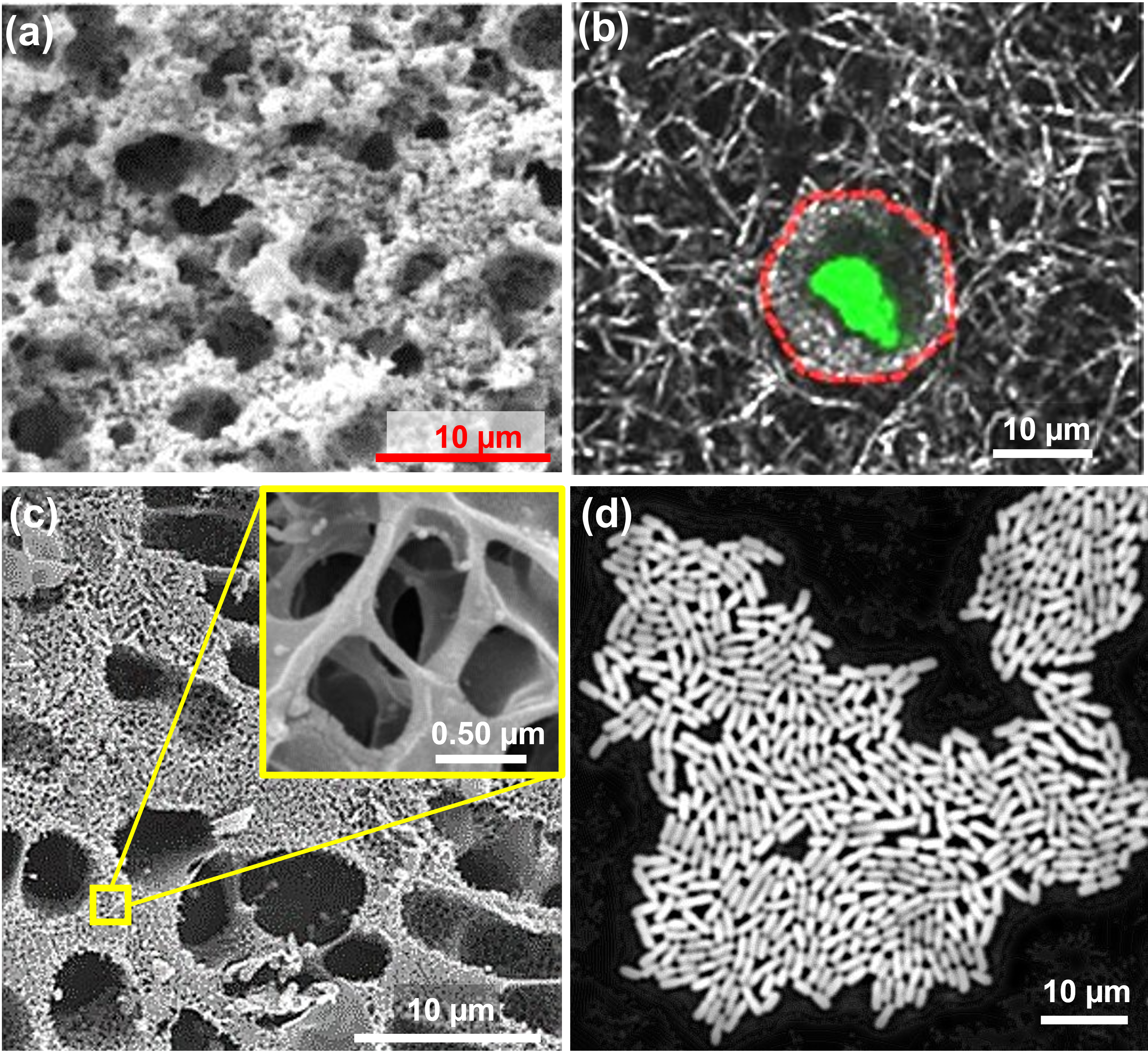}
  \caption{Porous structure of food and biological materials. (a) Scanning electron micrographs of Lighvan cheese reveal the porous structure, adapted with permission from \cite{aminifar2010}, copyright 2010 JohnWiley $\&$ Sons, Inc.. (b) Porosity of extracellular matrix affects cell migration and invasion of cancer tissues, adapted from \cite{carey2017}, CC BY 4.0. (c) Hierarchical porous structure of respiratory mucus revealed by cryo-SEM \cite{kirch2012}, Copyright (2012) National Academy of Sciences. (d) A growing bacterial colony transforming into a biofilm represents a poroelastic biomaterial.}
  \label{fig:food}
\end{figure*}

\paragraph*{Glacial weathering crust.} Glaciers cover about $10\%$ of the Earth's surface, but due to the current climatic change, the total surface area is shrinking at a significant rate as the glaciers melt. As a result, large quantities of frozen microorganisms ($\approx 2.9 \times 10^{22}$ cells per year) are entering the freshwater and marine ecosystems \cite{stevens2022spatially}. During the thawing process, microbes inhabit a slushy, saturated, porous matrix, characterized by a shallow ($<2\,\mathrm{m}$) ``weathering crust'' with the cryoconite (``cold dust'') holes piercing the continuous ice surface \cite{cook2016cryoconite}.
Cryoconites are active microbial hotspots where small ice crystals, fine silt (deposited by the wind), and diverse microbes coexist. Depending on the local temperature and humidity conditions, cyoconites may harbor millimeter-size aggregates of microbes and silt, held together by the ice crystals. Such aggregates represent labile saturated or unsaturated porous habitats \cite{hodson2010structure}.

\paragraph*{Food materials.} At the scale of microorganisms, food is a spatially structured complex soft material with internal energies of the order of the thermal energy, k$_{B}T$\cite{vilgis2016}. Spanning 1 nm to 100 $\mu$m in size, food microstructures can be classified as a single or multi-component self-organized porous systems consisting of polymer chains, proteins, surfactants, carbohydrates, and water and oil as selective solvents \cite{aminifar2010,skandamis2015}. In addition to the intrinsic structural complexity, food processing methods introduce microstructural changes that significantly affect the transport properties and shelf-life of the  food products \cite{ranjbaran2021}. The structural and functional effects of food processing have been captured using a porous media framework for a range of food products and processing parameters \cite{dadmohammadi2022}.

\paragraph*{Biological porous systems.} 
Across multiple scales, biological systems represent active, porous environments. Broadly speaking, bio-relevant porous tissues include (i) bones: Rigid porous structures filled with cells, extracellular matrix and biological fluids; (ii) pore networks, typically, the circulatory system: Hierarchical porous capillaries for blood, lymph and gas transport, ranging from $10 \,\mathrm{mm}$ aorta to the $5 \,\mathrm{\mu m}$ capillaries; and (iii) mucus linings: Deformable cross-linked polymer networks filled with viscoelastic fluids, with average pore sizes in the range of hundreds of nanometers \cite{bansil2013, bansil2018}. They form a semi-permeable viscoelastic coating on our eyes and the respiratory and  gastrointestinal tracts. A large proportion of mucus linings are inhabited by diverse microbes, such as the gut flora and the oral cavity.

\section*{Length scales in microbially relevant porous environments}

Depending on the physiological state, such biological porous environments can harbor disease-causing foreign parasites, including the African trypanosomes (\textit{Trypanosoma brucei}), which can actively swim through the bloodstream and the interstitial fluids \cite{kruger2018beyond}. Pathogenic bacteria are often found to swim in the lungs of patients suffering from cystic fibrosis \cite{zlosnik2014swimming}, while T-cells are versatile in their ability to navigate tissues such as liver, lymph nodes, and lungs \cite{rajakaruna2022liver}. Furthermore, modern medical technologies and biomedical diagnostics use active navigation strategies for disease detection and drug delivery \cite{gao2021biomedical, zhang2021dual, wu2018swarm, wan2020platelet}. Finally, the bacterial biofilm itself represents a living porous system. Embedded within an extracellular matrix composed of polymers and proteins, they act as active porous structures that evolve over time \cite{you2018geometry,you2019,carpio2019}, depending on local physical factors such as substrate curvature \cite{schamberger2023, langeslay2023}, structural anisotropy \cite {vlowery2019}, and substrate stiffness \cite{lin2020, sengupta2020, asp2022}. Interestingly, the porosity of the biofilm influences the growth rate of individual cells: cells occupying loosely packed regions of a colony show a higher rate of elongation compared to cells positioned in tightly packed regions of the biofilm \cite{witmann2023}.


Natural porous media are hierarchical in structure, with corresponding length scales spanning several orders of magnitude from sub-micron to millimeter length scales.
The small size of microorganisms makes them well-suited to live inside the pores. In the soil, quantitative studies show that the majority of bacteria are found in the micropores within soil aggregates \cite{han2021}.
Depending on the soil texture, the maximum diameter of the pores most frequently colonised by bacteria, has been estimated to be between $2.5$ and $9 \,\mu\mathrm{m}$, while no bacteria have been observed in pores smaller than $0.8 \,\mu\mathrm{m}$ in diameter \cite{ranjard2001quantitative}. 
The typical size of a \textit{Bacillus subtilis} bacterium is $4$--$5 \,\mathrm{\mu m}$ in length and $0.75$--$1\,\mathrm{\mu m}$ in diameter.
Experiments with microfluidic devices also show that the narrowest slit through which bacteria can pass is $0.75 \,\mu\mathrm{m}$ for \textit{B. subtilis} and $0.4 \,\mu\mathrm{m}$ for \textit{E. coli} ($1$--$2 \,\mathrm{\mu m}$ in length and $0.25$--$1\,\mathrm{\mu m}$ in diameter) \cite{mannik2009bacterial}. 
However, it still remains to be confirmed whether micropores act as hotspots for active microbes, or provide protective habitats under unfavorable conditions, or simply are passive dead-ends where cells are trapped \cite{totsche2018microaggregates}. Interestingly, in general, microbial colonies inside soil are tiny compared to lab-grown biofilms. For instance, \textit{B. subtilis}, which forms $\approx 10\,\mathrm{cm}$ large biofilms on nutrient-rich agarose culture plate, usually appears as small colonies of two to five cells \cite{grundmann2004spatial}. 

\begin{figure*}
  \centering
  \includegraphics[width=\textwidth]{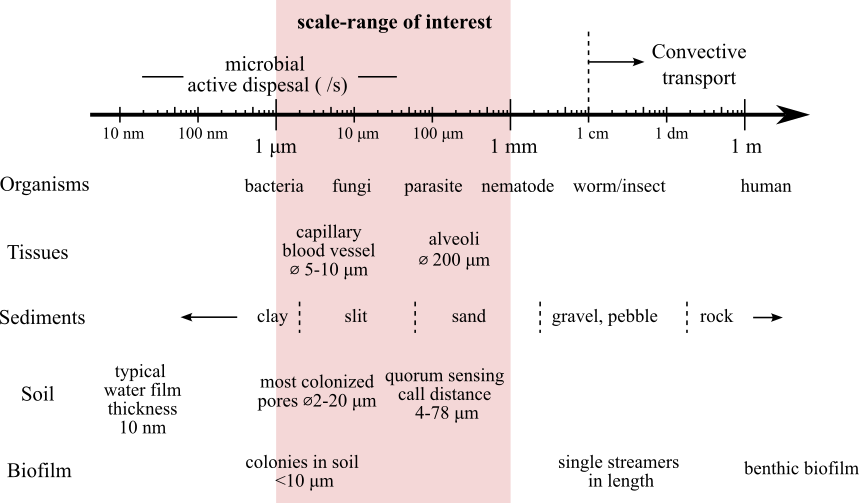}
  \caption{Characteristic length scales in microorganisms and their porous habitats. The shaded red region matches the typical microbial length scales, and represents the range of dimension covered in this review. Most of the numbers in the figure is also presented in the main text and from the following works \cite{dexter1988advances, ranjard2001quantitative, grundmann2004spatial, raynaud2014spatial, gantner2006situ, battin2016ecology}}
  \label{fig:scales}
\end{figure*}

Motile microbes can swim at a speed about one body length per second. 
The persistent swimming speed of \textit{Bacillus subtilis} can reach $ 20\,\mu$m/s \cite{najafi2019swimming}, 
while the parasite \textit{Trypanosoma brucei} ($8$ to $50 \,\mathrm{\mu m}$ long) swims at a average speed of $12\,\mathrm{\mu m/s}$  \cite{bargul2016species}.
Such a speed suggests that active dispersal of microbes occurs at the pore-scale as well as at scale of microbial body size, thus plays a key role in the accessing and colonisation of porous environments.
Both velocity and diffusion limit the range of cell-cell interactions.
Gantner et al studied the quorum sensing in the rhizosphere, and found that the effective ``call distance'' was between $4$ and $78 \,\mathrm{\mu m}$ depending on the environment \cite{gantner2006situ}.
Raynaud and Nunan analyzed microbial colonization patterns based on microscopic images of several hundred thin sections of soil. They observed an average cell-to-cell distance of about $12.5 \,\mathrm{\mu m}$ in the soil matrix \cite{raynaud2014spatial}.

\section*{Active microbes in dynamic porous confinements}
\begin{figure}
  \centering
  \includegraphics[width=0.7\columnwidth]{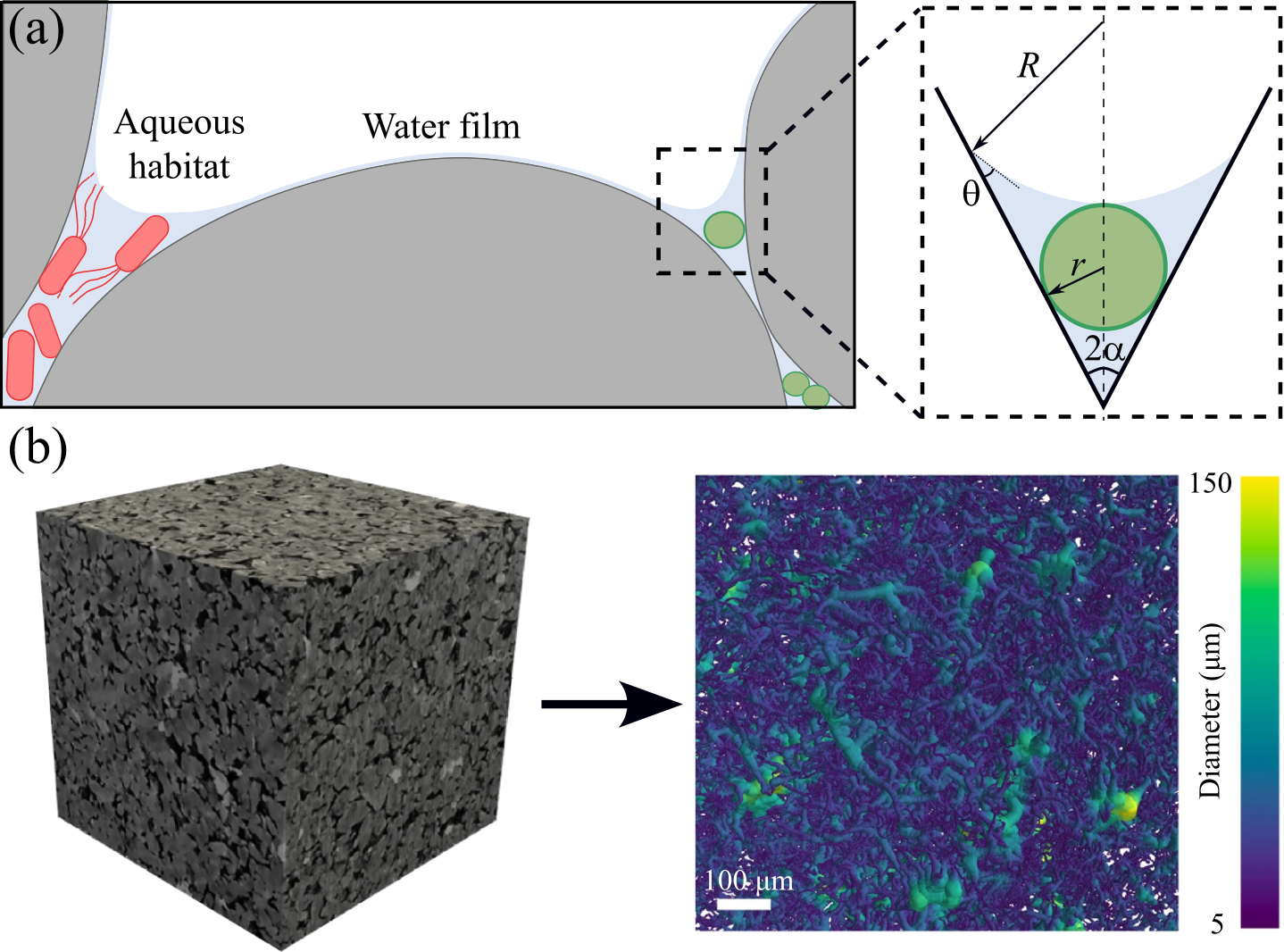}
  \caption{Porous environments impose a strong confinement on microorganisms. (a) Schematic representation of water distribution in a porous unsaturated environment, e.g., soil. The size of the confined aqueous zone, shown here in the corner, is determined by the contact angle and the angle between the solid surfaces. Depending on the size of the aqueous zone, the resident microbe could be completely or partially entrapped.
  (b) Pore network representation of a rock sample ($2.25\, \mathrm{mm}$ side length) derived from an X-ray microtomography, color-coded by local diameter \cite{neumann2021high}, CC BY 4.0.
  }
  \label{fig:confine}
\end{figure}

Microbes growing in porous environments are active systems where an interplay of confinement-induced physical and chemical constraints determine microbial behaviour, physiology and self-organization \cite{araujo2023}.
Porous surfaces can modulate the transport of microroganisms through steric, electrostatic, chemical and/or hydrodynamic interactions, each with a characteristic interaction range \cite{conrad2018confined}. 
Furthermore, transport processes in porous environments are mediated by the pore-scale network and topology \cite{hazas2022}. While random networks in most natural porous media are not perfectly connected (as shown in Fig.~\ref{fig:confine}(b)), the connectivity decreases further as the hierarchical length scale increases. This results in the formation of ``dead ends'' that can only be reached by diffusion or active dispersal of microorganisms. 

It should also be noted that most parts of the soil and subsurface are unsaturated with most of the water accumulates in the nooks and corners, as well as the micro- and mesopores.
It is the network of these aqueous elements, rather than the entire void, that limits both nutrient diffusion and microbial movement (Fig.~\ref{fig:confine}(a)) \cite{tecon2016bacterial}.
Specifically, Or et al estimated that in a typical soil with moderate moisture (soil matric potential $-30\, \mathrm{kPa}$), at a $60^o$ corner formed by two mineral surfaces, the aqueous element is just enough to submerge a single bacterium with a radius of $\approx 1\, \mathrm{\mu m}$ \cite{or2007physical}, but will limit the flagellar beating and hence the motility of microorganisms \cite{dechesne2010hydration, tecon2016bacterial}. 
The thin water film on a smooth mineral grain surface in such a soil ($<10\,\mathrm{nm}$) will also limit the diffusion of chemicals \cite{or2007physical}.
Both can lead to critical metabolic consequences for the microbes \cite{rivkina2000metabolic}.

Below is a brief overview of the different motility mechanisms that microbes use to navigate porous environments. More details on bacterial motility mechanisms can be found in \cite{wadhwa2022bacterial, thormann2022wrapped}.

\subsection*{Swimming}
Many microorganisms switch from a biofilm to a planktonic life form thanks to the flagella, which enable the cells to swim in aqueous environments after detachment from the mother colonies. Flagella are helical appendages, typically $5$ to $10 \,\mathrm{\mu m}$ long and $20 \,\mathrm{nm}$ in diameter, that are driven by molecular motors at the expense of energy units (ATP) to propel the cell body forward \cite{yonekura2003complete, grognot2021}. 
Although metabolically costly \cite{schavemaker2022flagellar}, swimming not only allows microorganisms to navigate, but also plays an important role in microbial colonization under flow (discussed in the next section). 

Over long timescales, bacterial swimming can be modelled as diffusion, involving straight run events, interspaced with reorientation events that allow cells to navigate and sample their environment. There are several reorientation mechanisms, the most thoroughly studied being ``run-and-tumble'': by changing the rotational state of the flagellar filament, the bacterial body alternates the state between ``run'' (approximately $1$s) and ``tumble'' (approximately $0.1$s), exhibiting a random walk on a large time scale \cite{lauga2016bacterial}.
For a long time, the run-time was considered to be Poisson distributed, but a recent study shows that there is a large behavioral variability due to the fluctuation of the signal protein \cite{figueroa20203d}.  
Under certain assumptions, the coarse-grained fluctuating hydrodynamics of interacting ABPs and run-tumble-particles can be mapped onto each other and are therefore strictly equivalent \cite{cates2013active}.

Bacteria are able to move towards nutrient, oxygen and quorum sensing signals, and away from predator signals. Such directional movement guided by an external chemical field is called chemotaxis \cite{keegstra2022ecological}.
By moving around, bacteria sample the chemical concentration in the environment, and then, based on very limited information, bias their random walk along / against the chemical gradient by reducing tumbling \cite{mattingly2021escherichia}. Confinement, including crowding due to high cell density, can affect the sampling process and hence the chemotaxis \cite{grognot2021multiscale, colin2019chemotactic}.
Chemotactic cells, even with some degree of phenotypic variability, can self-organize into travelling bands \cite{fu2018spatial, bhattacharjee2022chemotactic}. 
Such travelling bands can sweep rapidly across complex geometries, e.g., porous media \cite{adler1966chemotaxis,budrene1995dynamics, bhattacharjee2021chemotactic}, and are therefore used by microbes as an efficient active dispersal strategy \cite{cremer2019chemotaxis, ni2020growth}.

\subsection*{Swarming}
Swarming is a second form of flagella-mediated motility, typically observed when planktonic microbes grow in nutrient-rich environments. Under these conditions, cells become multinucleated, elongate, grow large numbers of flagella (hyperflagellation), and secret bio-surfactants to march across surfaces in coordinated rafts \cite{kearns2010field} 
Tumble mode is suppressed during swarming \cite{partridge2020tumble} 
The water content of the medium is a key determinant of swarming: while too much water promotes swimming, too little water inhibits swarming. 
Wetting agents, such as secreted biosurfactants and lipopolysaccharide (LPS), often aid swarming motion by maintaining an optimal moisture content of the solid surface.
To the best of our knowledge, there is currently no data on the optimal water content and film thickness required for swarming. Reports by Zhang \textit{et al.} have estimated the liquid film to be approximately 10 times thicker at the edge of the swarm than on the virgin agar \cite{zhang2010upper}, while Wang \textit{et al.} have found that spraying water (approximately $3\,\mu\mathrm{L}$ per $\mathrm{cm^2}$ of agar surface) stimulated cells to swarm normally \cite{wang2005sensing}.
Although swarming has not been demonstrated in natural environments, a similar form of collective movement of \textit{B. subtilis} towards plant roots has recently been observed in transparent soil \cite{engelhardt2022novel}. This collective movement may be a mechanism to enhance colonisation of the plant hosts.


\subsection*{Twitching}
Twitching is the bacterial surface movement powered by extension and retraction of type IV pili. The pili extend, attach to the surface and retract. The cell body is slowly pulled forward with a twitchy appearance \cite{mattick2002type, craig2019type}. Twitching motility is exhibited by a wide range of bacteria, many of which are pathogenic.
It has been observed on both organic and inorganic surfaces, including agar gels, epithelial cells, plastics, glass, and metals \cite{merz2000pilus, mattick2002type} 
This form of motility requires a moist surface but not bulk water. Twitching cells share EPS and move in groups to a new colonization site in low water environments at a speed of around $1 \, \mathrm{\mu m/s}$ \cite{merz2000pilus, zhao2013psl, oliveira2016single}.
The overall dynamics of twitching is also Brownian, possibly arising due to a ``tug-of-war'' between the pili located at different points on the cell body, similar to the cargo transported by molecular motors on cytoskeletal filaments \cite{marathe2014bacterial}. Twitching cells can exhibit chemotaxis \cite{oliveira2016single}.
Twitching mobility also allows bacteria to move against the current and spread upstream (discussed in the next section). 

\subsection*{Gliding}
The term ``gliding'' generally refers to non-flagellar active surface motions. These are smooth and continuous, and are observed in three major groups of bacteria, the myxobacteria ($0.025$ to $0.1 \,\mathrm{\mu m/s}$), the cyanobacteria (velocities approaching $10 \,\mathrm{\mu m/s}$), and the Cytophaga-Flavobacterium group ($2$ to $4 \,\mathrm{\mu m/s}$) \cite{pfreundt2023, faluweki2023}. 
Several mechanisms have been proposed to explain gliding motility, including propulsion by contraction of type IV pili (social gliding), transport of macromolecules such as polysaccharides and movement of outer membrane components by protein complexes in the cytoplasmic membrane (adventurous gliding). However a clear mechanistic picture of how gliding motility is achieved remains to be established.

\subsection*{Growth-induced sliding}
Sliding is driven by the mechanical force that dividing cells exert on their neighbours, and facilitated by bacterial secretions that help to reduce friction and promote sliding.  
It is not an active form of movement, but is likely to play an important role in bacterial surface colonization, particularly during the early stages of biofilm development when the rate of colony expansion (due to sliding) is proportional to the colony size \cite{dell2018growing, you2018geometry, dhar2022self}. More recently, the dispersal of individual cells within a growing colony has been linked to their genealogical separation from the mother cell, revealing the genealogical organization of bacterial cells in a growing colony \cite{rani2023}. Growth experiments in natural and artificial confinements have shown that the biomechanical forces generated in the colony are large enough to alter the growth rates of cells in densely packed regions \cite{witmann2023}, and even to squeeze them through slits which are narrower than the cell body \cite{mannik2009bacterial}, suggesting a relevant mechanism by which microbes could penetrate narrow necks to colonize porous confinements.

\subsection*{Microbial motility across length scales}

Despite their small size, microorganisms can move over multiple length scales, and use a variety of motility mechanisms, individually or collectively. 
The swimming speed of bacteria is quite fast: \textit{E.coli} can swim at about $25 \,\mathrm{\mu m/s}$, or about $20$ body lengths per second. However, they often tumble to change direction, which results in a lower speed of propagation.
When tumbling is suppressed, they are significantly faster: Cremer et al reported that \textit{E. coli} can move about $30\,\mathrm{mm}$ by collective swimming driven by chemotaxis \cite{cremer2019chemotaxis}. 
Swarming cells are even faster, \textit{B. subtilis} can travel $30\,\mathrm{mm}$ in just $2.5$ hours \cite{kearns2010field}.
Interestingly, when \textit{E.coli} moves upstream in a tube, a few pioneer cells reach the distance of $13\,\mathrm{mm}$ in about $13$ minutes \cite{figueroa2020coli}.
This is about $17\,\mathrm{\mu m/s}$ against a Poiseuille flow with a maximum velocity of $80\,\mathrm{\mu m/s}$. 

Bacteria can also travel centimetres by non-flagellar motions such as twitching, gliding, and growth-induced sliding. However, this takes hours or even days.
Twitching \textit{P. aeruginosa} has been observed to move more than
$0.7\,\mathrm{mm}$ against a flow (velocity $2$ to $20\,\mathrm{mm/s}$) in $15$ hours \cite{kim2015colonization}.
Different strains of \textit{B. subtilis} on different substrates grown on different substrates have colony fronts ranging from very slow (about $0.05 \,\mathrm{mm/h}$ or $0.016 \,\mathrm{\mu m/s}$) \cite{grau2015duo} to moderately fast ($5 \,\mathrm{mm/h}$ or $1.6 \,\mathrm{\mu m/s}$) \cite{kinsinger2003rapid}. 

It is worth noting that a high speed of individuals does not necessarily lead to high speed of collective migration: At high density, cells are organized into an active nematic structure by steric interactions. If individual cells move too fast, they can become trapped in vortices (local defects).
In this case, a lower individual speed actually results in a higher overall migration speed. Meacock et al. have found that this is indeed the strategy used by the wild type \textit{P. aeruginosa} twitching on surface \cite{meacock2021bacteria}.

\subsection*{Active colonization strategies in porous confinements}

Before they can colonize a site, passively transported microbes have to settle. For example, they must break out of the convective flow fields and attach to a suitable surface \cite{wheeler2019not}.
Their retention is determined by several factors including cell size and shape, motility, surface properties etc. \cite{zheng2021implication}
In particular, cell motility is generally thought to increase retention in porous media \cite{liu2011idling, creppy2019effect}.
However, the precise mechanisms leading to such observations remain to be elucidated, particularly in the context of interactions and biophysical feedback at the cellular and pore scales. 

In recent years, the development of microfluidic techniques has enabled quantitative studies of microbial ecology \cite{rusconi2014microfluidics}. 
Microfluidic techniques offer flexible and modular built-in structures that mimick the features within porous media, such as confinement, structured geometry, semi-permeable chamber and controlled flow.
Another advantage of microfluidics is its compatibility with optical imaging techniques, which can easily achieve micron-scale resolution at a reasonable cost.
Inside a microfluidic chamber, we could use micro-PIV (particle image velocimetry) to measure the flow profile, track the movement of individual bacteria, identify the orientation of the cell body and even the state of the flagella. 
With such a wealth of detailed information and mathematical models, researchers are able to uncover more mechanisms.
A better understanding of microscale mechanisms leads to significant progress in understanding and modelling of bacteria transport and dispersal at the macroscale. The following is an overview of the processes relevant to microbial dispersion and transport in porous media

\subsubsection*{Shear-induced retention at surface}

Rusconi et al. recorded the concentration profile of bacteria (\textit{B. subtilis}) across the width of a microfluidic channel in a Poiseuille flow \cite{rusconi2014bacterial}. 
As the flow (and shear rate) was increased, they observed a depletion of the central part of the profile of motile cells: up to $70\%$ of cells were depleted from low-shear regions (channel center) and accumulated in high-shear regions (channel boundaries).
Non-motile cells, on the other hand, showed a uniform distribution.
Combined with mathematical models, they showed that such accumulation results from the competition between the cell alignment with the flow and the stochasticity in the swimming orientation.
In contrast to the exclusion that often occurs in colloid transport, the shear-induced ``trapping'' increases the residence time of bacteria near the surface, thereby facilitating surface adhesion.
With a constriction in the flow path, passive particles show no preference for either residence site, but motile bacteria show a preference for both.
A counterintuitive observation is the increased cell density after a funnel \cite{altshuler2013flow}.
Similarly, motile bacteria in flow also accumulate on the leeward side of cylindrical obstacles \cite{mino2018coli,lee2021influence}.
These residential locations becomes colonization sites afterwards \cite{secchi2020effect}. 

\subsubsection*{Active motility against flow}
At moderate flow rates, motile bacteria may migrate upstream rather than be washed downstream, due to (positive) rheotaxis. 
Bacterial rheotaxis is a purely physical pheonomenon resulting from an interplay of fluid shear, asymmetric body shape, flagellar rotation, and the surface interaction. 
The shear-induced torque orients the swimming direction of the microorganisms preferentially upstream \cite{mathijssen2019oscillatory}.
Figueroa-Morales et al. suggested a link between the rheotaxis and large scale ``super-contamination'' of biological channels, catheters or water resources \cite{figueroa2020coli} 
Rheotaxis can occur in bulk flow, induced by the chirality and geometrical features of bacteria \cite{marcos2012bacterial,jing2020chirality}.
When confined in narrow channels, experiments with artificial microswimmers show that shape asymmetry is no longer necessary \cite{dey2022oscillatory}.

In addition to swimming bacteria, sessile bacteria can also move against the current: In a flow environment, \textit{P. aeruginosa} orient themselves with the type IV pili pole pointing in the opposite direction of the flow, and then they twitch upstream \cite{kim2015colonization} 
This mechanism allows them to disperse in vasculature-like flow networks.
These observations suggest that the dependence of the bacterial orientations on fluid shear, which further influences the transport behaviour within porous media.

\subsubsection*{``Catch-bonds'' under shear stress}
Bacteria have unique feedback mechanisms to counteract biomechanical cues, such as catch-bonds. Thomas et al. reported that the attachment strength of \textit{E. coli} to monomannose coated surfaces depended on the external shear stress. Increasing the shear stress increased the accumulation of cells on surfaces by up to two orders of magnitude, whereas decreasing the shear stress led to cell detachment. \cite{thomas2002bacterial,thomas2004shear,thomas2008biophysics}. 
Such attachment occurs via ``catch-bonds'' mediated by the type 1 fimbrial adhesive subunit, FimH, a force-responsive mechanism. At low shear rates, cells form only the normal ``slip-bonds''. At increased shear rates most cells detached, but some stick firmly due to the formation of ``catch-bonds'' induced by the high shear.
Although ``catch-bonds'' are only reported on the \textit{E. coli} adhesion, FimH is the most common type of bacterial adhesin known, it is possible that other taxa have similar mechanisms.

\subsubsection*{Environmental engineering by microorganisms}
Bacteria are able to secret numerous polymeric substances to modify their local environment, as means to protect themselves from harmful environmental conditions and to support the biofilm life form \cite{flemming2019bacteria}.
These substances are generally referred to as extracellular polymeric substances (EPSs), and typical EPS includes polysaccharides, proteins, short-chain DNAs, biosurfactants and lipids.
EPSs tune the surface properties of microbial cells. This could facilitate transport or retention of the cells depending on the type of the substance \cite{zhong2017transport}. 
Interestingly, one of the confirmed triggers for EPS production is fluid shear \cite{weaver2012fluid,rodesney2017mechanosensing}.
Prior to adhesion, bacteria produce EPS near the surface. Although still in aqueous solution, EPS form a weak network through entanglements and ion-mediated intermolecular associations \cite{ganesan2016associative}, which becomes a precursor of biofilm.
The Psl exopolysaccharide produced by \textit{P. aeruginosa} not only acts as a ``molecular glue'' and promotes surface attachment \cite{ma2006analysis}, but also acts as an autochemotactic trail that accumulates cells and initiates the colonization \cite{zhao2013psl}.

In saturated, nutrient-rich regions, such as the riverbeds, microorganisms grow well-developed and highly differentiated biofilms. 
In addition to the surface-attached, flat biofilms, there is a special type of filamentous biofilm that follows the streamline, called ``streamers''. Streamers can cause bioclogging and alter the hydraulic properties of the porous media \cite{drescher2013biofilm, battin2016ecology} 

Even without the production of biochemicals, the swimming motion itself can already modify the environment.
Friedlander et al. have reported that when the bacteria swim over a Cassie-Baxter wetted surface (air bubbles trapped in between surface textures), the beating flagella will pump energy into the system, break the Cassie-Baxter wetting state and changing it into a Wenzel state (air bubbles released), thereby increasing the surface area available for adhesion \cite{friedlander2013bacterial}.

\subsection*{ Active particles on a random landscape}
A Brownian particle in an unconfined space move diffusely. If the confinement is strong, i.e., when gas molecules diffuse through very narrow pores, the size of which is comparable to or smaller than the mean free path of the gas, their migration is described by Knudsen diffusion, which is one of the is one of the models of subdiffusion. 
Microbial migration within porous media falls into a similar range, where the bacterial ``mean free path'' (mean run length) is of the same order of magnitude as the typical pore diameter \cite{duffy1995random}.
Note that the ``run length'' here is not limited to the swimming bacteria, but also refers to the persistent length for twitching bacteria on the surface.

Due to their activity, unconfined bacteria migrate superdiffusively in the short range, and diffusively in the long range.
In simulations they are often represented as active particles, such as active Brownian particles (ABPs) \cite{zeitz2017active}, run-tumble particles (RTPs) \cite{reichhardt2014active}. 
The overall dynamics depends on several parameters in the system, including the activity of the particles, the reorientation frequency, the density of obstacles, etc. \cite{chepizhko2013diffusion,reichhardt2014active,morin2017diffusion,zeitz2017active, spagnolie2023swimming}
Typically, when the density of obstacles is high, and the frequency of reorientation is low (corresponding to long run lengths), the particles will enter a sub-diffusion regime and even becomme trapped. 

Different strategies are required to navigate efficiently in an open space or a disordered environment.
Volpe et al have suggested that a diffusive search is more efficient than a ballistic search when navigating a complex topography with boundaries, barriers and obstacles \cite{volpe2017topography}.
For the run-tumble particles, several studies using different models have predicted an optimized tumbling probability in order to maximize the diffusivity in disordered environments \cite{reichhardt2014active,licata2016diffusion,bertrand2018optimized,irani2022dynamics}.  
This is qualitatively explained by experimental observations: increased tumbling frequency also increases the migration efficiency of swimming bacteria inside agarose \cite{wolfe1989migration}. 

More recently, Bhattacharjee et al. have developed a transparent porous medium that allows three-dimensional tracking of bacterial movement within it \cite{bhattacharjee2019bacterial,bhattacharjee2019confinement,perez2021impact}.
They have observed that due to the pore-scale confinement, the bacteria exhibit a ``hop-and-trap'' motion rather than ``run-and-tumble'' motion. Cells are intermittently and transiently trapped and then hop to the next position.
While ``hopping'' is determined by pore-scale confinement, and is independent of cellular activity; ``trapping'' is determined by the competition between pore-scale confinement and cellular activity.
Kurzthaler et al have reproduced the ``hop-and-trap'' behaviour in Brownian dynamics simulations of active stiff polymers undergoing run-reverse motion in porous media. They have also proposed a geometric criterion for the optimal spreading when the ballistic ``run length'' of the bacteria is comparable to the ``longest straight path'' in the porous medium \cite{kurzthaler2021geometric}.
Dehkharghani et al identified the key geometric parameter controlling cell transport is the ``effective mean free path'', which is calculated as the average distance for randomly sampled, theoretical straight paths between solid surfaces \cite{dehkharghani2023self}. 

\subsection*{Microbial motion inside porous media: Zooming into the pore scale}

\begin{figure*}
  \centering
  \includegraphics[width=\textwidth]{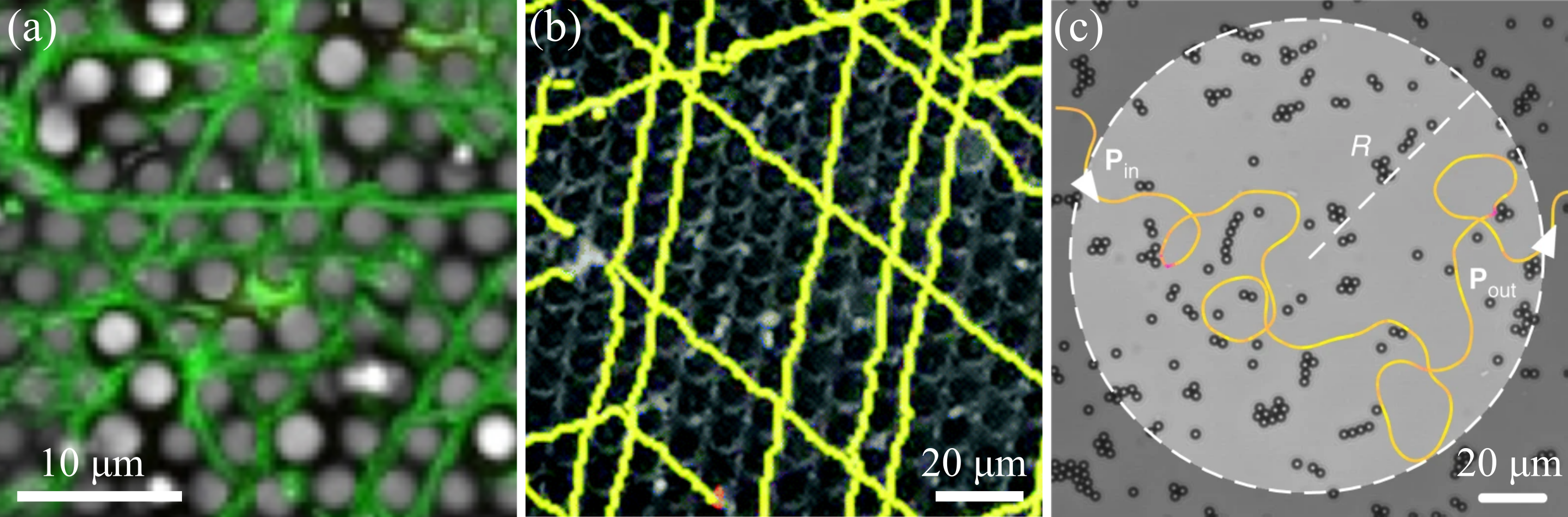}
  \caption{Length scaling matching between cells and porous confinements alter the dynamics of microbial locomotion. (a) Twitching \textit{P. aeruginosa} on a surface patterned with arrays of hemispheres follow the crystal axis (reproduced with permission from Reference \cite{chang2018surface}, copyright 2017 American Chemical Society). (b) Swimming \textit{E. coli} inside a crystal (adapted from Reference \cite{brown2016swimming}, CC BY 3.0). (c) Small obstacles at a certain density can scatter swimming \textit{E. coli} forward (adapted from Reference \cite{makarchuk2019enhanced}, CC BY 4.0)}
  \label{fig:pattern}
\end{figure*}

Microorganisms are in constant interaction with their local environment while in motion. For example, the hydrodynamic and steric interactions of the swimming bacteria with the pore surface will easily alter their trajectories as discussed in previous sections. Twitching bacteria need to attach their pili to a solid, so the landscape of the environment will also change their dynamics. 
Figure~\ref{fig:pattern}(a) and (b) show similar dynamics in two different systems: one is  \textit{P. aeruginosa} twitching on a surface patterned with arrays of colloidal hemispheres with an inter-sphere spacing of $4\,\mathrm{\mu m}$ \cite{chang2018surface}, the other is \textit{E. coli} swimming inside a colloidal crystal with the inter-colloid spacing of $\approx 7\,\mathrm{\mu m}$ \cite{brown2016swimming}.
Both of them show preferential motion along the crystal axes. 
Note that such induced persistent motion is only observed when the length scale of the bacteria motion and the local environment are matched. 
When the same \textit{P. aeruginosa} twitches on patterned surface with larger or smaller inter-sphere spacing \cite{chang2018surface}, or when swimming bodies with a different rotational dynamics (\textit{E. coli} with shorter flagella and Janus swimmers) swim inside the same colloidal crystal \cite{brown2016swimming}, the movement no longer follows the crystal axes and the mean squared displacement (MSD) in the same time interval decreases.

Randomly placed obstacles can also promote the spread of swimming bacteria. Makarchuk et al have reported that small obstacles placed at the right density significantly increase the spread of swimming \textit{E. coli} through individual forward scattering events. \cite{makarchuk2019enhanced} (Fig.~\ref{fig:pattern}(c)).

\subsection*{Confined microorganisms: escape or settle down?}
Although bacteria are often found in pores comparable in size to cells, it remains to be confirmed whether cells have actively chosen to occupy these spaces or have been passively deposited. In addition, the impact of pore size confinement on biological activity and metabolism remains to be elucidated.
Existing reports support the hypothesis that bacteria are able to sense the contact with a surface \cite{o2016sensational}, and even discriminate between different surface properties \cite{friedlander2013bacterial, yang2022bacterial}.
Geng et al reported that mere contact with a surface can reduce the level of cellular respiration in certain strains of \textit{E. coli} \cite{geng2014bacteria}.
One of the mechanisms responsible for surface sensing is the ``flagellar dynamometer'': upon contact with a surface, the reduced rotation of the flagellar filament initiates a signalling pathway that triggers a change in bacterial state to swarming or biofilm formation \cite{o2016sensational}.
Type IV pili have also been shown to play an important role in surface sensing.  

One would naturally expect that, geometric confinement would also induce changes in microbial metabolism and dynamics, directly via mechanical, and indirectly via chemical cues. 
Chu et al demonstrated that the mechanical stress of bacteria filling a microfluidic chamber induces biofilm formation \cite{chu2018self}.
In some other cases, confinement ($2$ to $3$ body sizes) triggers the escape response of bacteria \cite{lynch2022transitioning}, and acts as a critical cue for \textit{Vibrio fischeri} to colonize their animal host.
The acceleration of swimming bacteria under confinement is also seen in other systems \cite{yin2022escaping} and is explained as a hydrodynamic effect.  
Geometric confinement can also increase the local concentration of chemical signals, inducing quorum sensing that normally occurs only in high cell densities \cite{carnes2010confinement}.

Very strict confinement, however, limits the mobility of the bacteria.
Experiments show that the limit of flagella-mediated motility is reached when the cell is in a slit of less than $\approx 30\%$ wider than the body width \cite{mannik2009bacterial}.
Bacteria (\textit{E. coli} and \textit{B. subtilis} in this study) still have the ability to penetrate such a slit with growth-induced sliding motility \cite{mannik2009bacterial}.
Other escape mechanisms exist: \textit{Shewanella putrefaciens} can escape from a narrow pore with a screw-like movement \cite{kuhn2017bacteria}.


\section*{Summary and perspectives}

Building on the recent insights gained at the interface of microbial ecology and porous media, in this article we have we reviewed key parameters--both physical and biological--which have advanced our understanding of the active processes and feedback that govern microbial behaviour and physiology under diverse porous settings. How physical constraints, in conjugation with active biophysical mechanisms, mediate microbial fitness and survival is an emerging field of study. Specifically, the ability of microbes to shape their local environments feeds back into their behaviour and physiology, triggering interactions at across individual, population and community scales. More broadly, these new insights shed light on emerging microbiome structures, assemblages and functions under porous conditions (Figure \ref{fig:summary}). The active strategies covered in this review should provide the fundamental concepts for dissecting microbial interactions with their neighbours and porous boundaries under ecologically-relevant conditions. Taking cues from the nature, systematic mechanistic studies can be designed to uncover the microbial feedback loops and the next generation of bioremediation and health management suites \cite{borer2021, sengupta2020, philippot2023, jansson2023, bansil2018, wu2023mucin}.

As cross-disciplinary multi-scale experimental approaches become the norm, data on microbial behavior and physiology, in relation to the micro-environmental properties, are beginning to be widely accessible. Our ability to zoom into the micro-scale dynamics is expected to reveal how the local environment -- both spatially and temporally -- shapes the non-equilibrium dynamics underpinning emergent microbial interaction networks. Importantly, new insights into how microbes and microbiomes shape local constraints, have clearly highlighted the need to develop an ensemble approaches to study microbe-matrix interfaces. Machine learning approaches including deep neural networks for feature recognition and tracking; and recurrent nets and random forests for time series analysis \cite{cichos2020} could be a timely incorporation to develop a comprehensive understanding. Unravelling microbial interactions and feedback in porous environments will have far-reaching implications for biomedicine, food sciences, sustainability, and environmental biotechnology, alongside fundamental scientific discoveries. To help set the tone for the way forward, we conclude the review with a selection of key open questions that await exploration:

\begin{figure*}
  \centering
  \includegraphics[width=.6\textwidth]{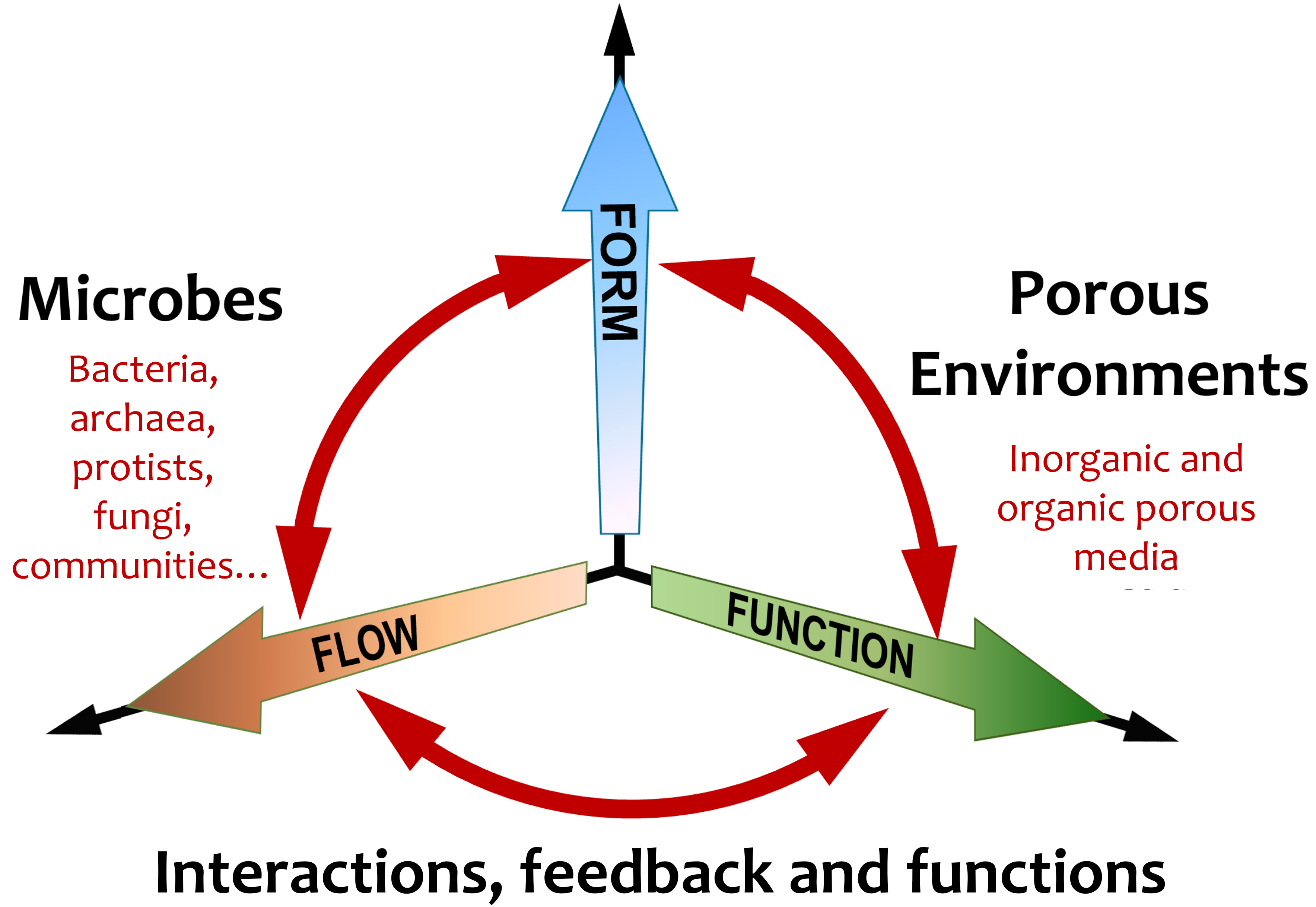}
  \caption{Understanding microbial feedback and interactions in porous environments offers fundamental insights into the emergence of active behavioural and physiological strategies, relevant in a wide range of health and ecosystem applications. In the future, mechanistic studies of community-scale interactions -- involving diverse microbial species in dynamic porous environments -- could be designed to advance the next generation of biomedical and environmental solutions.}
  \label{fig:summary}
\end{figure*}

\begin{itemize}
  \item \textbf{Role of noise.} Microbial behaviour and interactions in porous environments are modelled as active Brownian particles, taking into account the mean values of key parameters such as cell size, division rates, motility, rotational diffusion etc. Natural systems, however, are \textit{noisy} in a statistical sense \cite{dhar2022self, figueroa20203d}. Although the role of noise has recently been discussed in the context of biofilm development \cite{dhar2022self}, its corresponding role in mediating active interactions and feedback under porous constraints remains unexplored for both geological and biomedical settings.
  
  \item \textbf{Life form shifts.} Microbes confined in porous environments dynamically switch between motile and sessile life forms. While bacterial life forms shifts have been extensively investigated in the context of biofilm formation and their dispersal, it is unclear if, how, and to what extent strict confinement conditions affect the ability of species to undergo such shifts. Strict confinement induces secondary physical effects, such as the enhancement of local gradients (chemical, pH, salinity etc.), suggesting potential roles of confounding factors that could mediate life form shifts in pore-confined microbes. 
  
  \item \textbf{From microbes to microbiome.} Advances in the design and engineering of synthetic microbiomes, together with the progress in characterization toolkits (physiological, metabolic and molecular), have driven investigations of community-scale microbial network structures, functions and resilience \cite{jansson2023}. Overlaying such networks with ecologically relevant models of microbial habitats could reveal functional interactions among species, and advance predictive modelling of complex metabolic landscapes with relevant implications for the microbial structures and functions in natural environments \cite{borer2021}.
  
  
 \item \textbf{Genealogical organization.} Recent studies have revealed the emergence of distinct self-similar genealogical enclaves within growing bacterial colonies, pointing to the competing roles of growth-induced forces and adhesive interactions in shaping the genealogical distance of cells to their kith and kin \cite{rani2023, meacock2023}. Leveraging tracking-based approaches to follow cell genealogy in porous environments will lead to fundamental insights into the organization of genealogical domains in the presence of physical obstacles, with varying cell-surface adhesive interactions. 

\item \textbf{Self-assembly of complex biological molecules.} Porous substrates enhance local molecular concentrations and nucleate highly ordered material phases of biological significance \cite{vamsee2023}, for example through the selective absorption of microbe-generated lyotropic biosurfactants. The self-assembly of lyotropic liquid crystalline phases can be fine-tuned by controlling the flow through the porous structures, via geometry, topology, or local surface properties \cite{sengupta2013}, triggering biologically relevant phase transitions and biochemical feedback loops within the microbe-inhabited porous environments. This yet-to-be-explored biochemical self-organization could provide key insights into the transformation of simple molecules such as nucleotides into complex biological molecules, such as long RNA chains. 
 
 \item \textbf{Subsurface Hydrogen storage.} One of the critical challenges associated with the subsurface storage of H$_{2}$ -- a key prospect for the next generation of renewable energy sources -- is the interaction of H$_{2}$ with microbes and its consequences. H$_{2}$ metabolizing bacteria can form gaseous by-products and biofilms, which can adversely affect the porosity and permeability through bioclogging. Furthermore, the production of biosurfactants could alter the wettability of the impacting transport of H$_{2}$ gas. Bottom-up studies aimed to address these open questions are currently lacking: they will be crucial to advance the mechanistic understanding of emerging scenarios arising from microbial interactions and feedback arising across different physical scales.

\end{itemize}

\section*{Acknowledgments}
 We gratefully acknowledge the support from the Institute for Advanced Studies, University of Luxembourg (AUDACITY Grant: IAS-21/CAMEOS to A.S.) and a Marie Skłodowska-Curie Actions Individual Fellowship (Project 101110587-MINIMA to C.J.). A.S. thanks Luxembourg National Research Fund for the ATTRACT Investigator Grant (A17/MS/ 11572821/MBRACE) and a CORE Grant (C19/MS/13719464/TOPOFLUME/Sengupta) for supporting this work. 

\section*{Author contributions}
Conceptualization, planning, administration, and supervision: A.S. Literature survey and analysis: C.J. and A.S. Writing: C.J. (initial draft) and A.S. (revision and final draft). \\ \\




\newpage

\bibliography{Navigation-sciadv-arXiv}

\begin{thebibliography}{100}

\bibitem{flemming2019bacteria}
H.-C. Flemming, S.~Wuertz, {\it Nature Reviews Microbiology\/} {\bf 17}, 247 (2019).

\bibitem{bar2018biomass}
Y.~M. Bar-On, R.~Phillips, R.~Milo, {\it Proceedings of the National Academy of Sciences\/} {\bf 115}, 6506 (2018).

\bibitem{philippot2023}
L.~Philippot, C.~Chenu, A.~Kappler, M.~C. Rillig, N.~Fierer, {\it Nature Reviews Microbiology\/}  (2023).

\bibitem{ebrahimi2014}
A.~Ebrahimi, D.~Or, {\it Water Resources Research\/} {\bf 50}, 7406 (2014).

\bibitem{dang2016}
H.~Dang, C.~R. Lovell, {\it Microbiology and molecular biology reviews\/} {\bf 80}, 91 (2016).

\bibitem{kirch2012}
J.~Kirch, {\it et~al.\/}, {\it Proceedings of the National Academy of Sciences\/} {\bf 109}, 18355 (2012).

\bibitem{bansil2013}
R.~Bansil, J.~Celli, J.~Hardcastle, B.~Turner, {\it Frontiers in immunology\/} {\bf 4}, 310 (2013).

\bibitem{bansil2018}
R.~Bansil, B.~Turner, {\it Advanced drug delivery reviews\/} {\bf 124}, 3 (2018).

\bibitem{wu2023mucin}
C.~M. Wu, {\it et~al.\/}, {\it npj Biofilms and Microbiomes\/} {\bf 9}, 11 (2023).

\bibitem{aminifar2010}
M.~Aminifar, M.~Hamedi, Z.~Emam-Djomeh, A.~Mehdinia, {\it Journal of Texture Studies\/} {\bf 41}, 579 (2010).

\bibitem{ranjbaran2021}
M.~Ranjbaran, B.~Carciofi, A.~Datta, {\it Comprehensive Reviews in Food Science and Food Safety\/} {\bf 20}, 4213 (2021).

\bibitem{dadmohammadi2022}
Y.~Dadmohammadi, A.~Datta, {\it Food Reviews International\/} {\bf 38}, 953 (2022).

\bibitem{kiorboe2002}
T.~Kiørboe, H.~Grossart, H.~Ploug, K.~Tang, {\it Applied and Environmental Microbiology\/} {\bf 68}, 3996 (2002).

\bibitem{nguyen2017}
T.~Nguyen, F.~Tang, F.~Maggi, {\it Journal of Geophysical Research: Earth Surface\/} {\bf 122}, 1794 (2017).

\bibitem{lawrence2023}
T.~Lawrence, {\it et~al.\/}, {\it Frontiers in Earth Science\/} {\bf 11}, 1264953 (2023).

\bibitem{borer2023}
B.~Borer, I.~Zhang, A.~Baker, G.~O'Toole, A.~Babbin, {\it PNAS Nexus\/} {\bf 2}, pgac311 (2023).

\bibitem{kondakova2013study}
G.~Kondakova, N.~Verkhovtseva, S.~Ostroumov, {\it Moscow University biological sciences bulletin\/} {\bf 68}, 119 (2013).

\bibitem{heinemann2021}
N.~Heinemann, {\it et~al.\/}, {\it Energy and Environ. Sci.\/} {\bf 14}, 853–864 (2021).

\bibitem{muhammed2022}
N.~S. Muhammed, {\it et~al.\/}, {\it Energy Rep.\/} {\bf 8}, 461–499 (2022).

\bibitem{thaysen2021}
E.~M. Thaysen, {\it et~al.\/}, {\it Renew. Sustain. Energy Rev.\/} {\bf 151}, 111481 (2021).

\bibitem{liu2023}
N.~Liu, A.~Kovscek, M.~Fernø, N.~Dopffel, {\it Frontiers in Energy Research\/} {\bf 11}, 1124621 (2023).

\bibitem{eddaoui2021}
N.~Eddaoui, M.~Panfilov, L.~Ganzer, B.~Hagemann, {\it Transp. Porous Media\/} {\bf 139}, 89–108 (2021).

\bibitem{liu2019}
N.~Liu, {\it et~al.\/}, {\it J. Industrial Microbiol. Biotechnol.\/} {\bf 46}, 855–868 (2019).

\bibitem{santha2019}
N.~Santha, P.~Cubillas, C.~Greenwell, {\it IOR 2019: 20th European Symposium on Improved Oil Recovery\/} {\bf 2019}, 1 (2019).

\bibitem{pan2022}
B.~Pan, {\it et~al.\/}, {\it Energy and Fuels\/} {\bf 36}, 4268–4275 (2022).

\bibitem{liu2023pore}
N.~Liu, A.~R. Kovscek, M.~A. Fern{\o}, N.~Dopffel, {\it Frontiers in Energy Research\/} {\bf 11}, 1124621 (2023).

\bibitem{power2007biologically}
I.~M. Power, S.~A. Wilson, J.~M. Thom, G.~M. Dipple, G.~Southam, {\it Geochemical Transactions\/} {\bf 8}, 1 (2007).

\bibitem{hodson2010structure}
A.~Hodson, {\it et~al.\/}, {\it Journal of Glaciology\/} {\bf 56}, 349 (2010).

\bibitem{conrad2018confined}
J.~C. Conrad, R.~Poling-Skutvik, {\it Annual review of chemical and biomolecular engineering\/} {\bf 9}, 175 (2018).

\bibitem{rhodeland2020}
B.~Rhodeland, K.~Hoeger, T.~Ursell, {\it Journal of the Royal Society Interface\/} {\bf 17}, 20200147 (2020).

\bibitem{tecon2017biophysical}
R.~Tecon, D.~Or, {\it FEMS microbiology reviews\/} {\bf 41}, 599 (2017).

\bibitem{scheidweiler2020}
D.~Scheidweiler, F.~Miele, H.~Peter, T.~J. Battin, P.~de~Anna, {\it Journal of The Royal Society Interface\/} {\bf 17}, 20200046 (2020).

\bibitem{balseiro-romero2022}
M.~Balseiro-Romero, {\it et~al.\/}, {\it Environmental Science and Technology\/} {\bf 56}, 13975 (2022).

\bibitem{wisnoski2023scaling}
N.~I. Wisnoski, J.~T. Lennon, {\it Trends in Microbiology\/} {\bf 31}, 242 (2023).

\bibitem{wu2023}
Y.~Wu, {\it et~al.\/}, {\it Nature Communications\/} {\bf 14}, 4226 (2023).

\bibitem{mcdougald2012}
D.~McDougald, S.~A. Rice, N.~Barraud, P.~D. Steinberg, S.~Kjelleberg, {\it Nature Reviews Microbiology\/} {\bf 10}, 39 (2012).

\bibitem{smith2018}
H.~J. Smith, {\it et~al.\/}, {\it FEMS microbiology ecology\/} {\bf 94}, fiy191 (2018).

\bibitem{coyte2017microbial}
K.~Z. Coyte, H.~Tabuteau, E.~A. Gaffney, K.~R. Foster, W.~M. Durham, {\it Proceedings of the National Academy of Sciences\/} {\bf 114}, E161 (2017).

\bibitem{taubert2018}
M.~Taubert, {\it et~al.\/}, {\it Environmental microbiology\/} {\bf 20}, 369 (2018).

\bibitem{abiriga2022}
D.~Abiriga, A.~Jenkins, H.~Klempe, {\it Annals of Microbiology\/} {\bf 72}, 1 (2022).

\bibitem{soares2023}
A.~Soares, {\it et~al.\/}, {\it Microbiology\/} {\bf 169} (2023).

\bibitem{konig2020physical}
S.~K{\"o}nig, H.-J. Vogel, H.~Harms, A.~Worrich, {\it Frontiers in Ecology and Evolution\/} {\bf 8}, 53 (2020).

\bibitem{vos2013micro}
M.~Vos, A.~B. Wolf, S.~J. Jennings, G.~A. Kowalchuk, {\it FEMS microbiology reviews\/} {\bf 37}, 936 (2013).

\bibitem{volk2016}
E.~Volk, S.~C. Iden, A.~Furman, W.~Durner, R.~Rosenzweig, {\it Water Resources Research\/} {\bf 52}, 5813 (2016).

\bibitem{rabbi2020microbial}
S.~M. Rabbi, B.~Minasny, A.~B. McBratney, I.~M. Young, {\it Geoderma\/} {\bf 360}, 114033 (2020).

\bibitem{totsche2018microaggregates}
K.~U. Totsche, {\it et~al.\/}, {\it Journal of Plant Nutrition and Soil Science\/} {\bf 181}, 104 (2018).

\bibitem{escudero2023}
C.~Escudero, R.~Amils, {\it Frontiers in Astronomy and Space Sciences\/} {\bf 10}, 1203845 (2023).

\bibitem{escudero2018active}
C.~Escudero, M.~Vera, M.~Oggerin, R.~Amils, {\it Scientific reports\/} {\bf 8}, 1 (2018).

\bibitem{tarkowski2022towards}
R.~Tarkowski, B.~Uliasz-Misiak, {\it Renewable and Sustainable Energy Reviews\/} {\bf 162}, 112451 (2022).

\bibitem{harris2007situ}
S.~H. Harris, R.~L. Smith, J.~M. Suflita, {\it FEMS microbiology ecology\/} {\bf 60}, 220 (2007).

\bibitem{carey2017}
S.~Carey, K.~Martin, C.~Reinhart-King, {\it Scientific Reports\/} {\bf 7}, 42088 (2017).

\bibitem{stevens2022spatially}
I.~T. Stevens, {\it et~al.\/}, {\it Communications Earth \& Environment\/} {\bf 3}, 1 (2022).

\bibitem{cook2016cryoconite}
J.~Cook, A.~Edwards, N.~Takeuchi, T.~Irvine-Fynn, {\it Progress in Physical Geography\/} {\bf 40}, 66 (2016).

\bibitem{vilgis2016}
T.~Vilgis, H.~J. Limbach, {\it Journal of Physics D: Applied Physics\/} {\bf 49}, 110401 (2016).

\bibitem{skandamis2015}
P.~N. Skandamis, S.~Jeanson, {\it Frontiers in Microbiology\/} {\bf 6}, 1178 (2015).

\bibitem{kruger2018beyond}
T.~Kr{\"u}ger, S.~Schuster, M.~Engstler, {\it Trends in parasitology\/} {\bf 34}, 1056 (2018).

\bibitem{zlosnik2014swimming}
J.~E. Zlosnik, {\it et~al.\/}, {\it PLoS One\/} {\bf 9}, e106428 (2014).

\bibitem{rajakaruna2022liver}
H.~Rajakaruna, J.~H. O’Connor, I.~A. Cockburn, V.~V. Ganusov, {\it The Journal of Immunology\/} {\bf 208}, 1292 (2022).

\bibitem{gao2021biomedical}
C.~Gao, {\it et~al.\/}, {\it Advanced Materials\/} {\bf 33}, 2000512 (2021).

\bibitem{zhang2021dual}
H.~Zhang, {\it et~al.\/}, {\it Science Robotics\/} {\bf 6}, eaaz9519 (2021).

\bibitem{wu2018swarm}
Z.~Wu, {\it et~al.\/}, {\it Science advances\/} {\bf 4}, eaat4388 (2018).

\bibitem{wan2020platelet}
M.~Wan, {\it et~al.\/}, {\it Science advances\/} {\bf 6}, eaaz9014 (2020).

\bibitem{you2018geometry}
Z.~You, D.~J. Pearce, A.~Sengupta, L.~Giomi, {\it Physical Review X\/} {\bf 8}, 031065 (2018).

\bibitem{you2019}
Z.~You, D.~J. Pearce, A.~Sengupta, L.~Giomi, {\it Physical Review Letters\/} {\bf 123}, 178001 (2019).

\bibitem{carpio2019}
A.~Carpio, E.~Cebrián, P.~Vidal, {\it International Journal of Non-Linear Mechanics\/} {\bf 109}, 1 (2019).

\bibitem{schamberger2023}
B.~Schamberger, a.~et, {\it Advanced Materials\/} {\bf 35}, 2206110 (2023).

\bibitem{langeslay2023}
B.~Langeslay, G.~Juarez, {\it Soft Matter\/} {\bf 19}, 3605 (2023).

\bibitem{vlowery2019}
N.~Vallespir~Lowery, T.~Ursell, {\it Proceedings of the National Academy of Sciences\/} {\bf 116}, 379 (2019).

\bibitem{lin2020}
H.~Lin, M.~T. Suleiman, D.~G. Brown, {\it Soils and Foundations\/} {\bf 60}, 944 (2020).

\bibitem{sengupta2020}
A.~Sengupta, {\it Frontiers in Physics\/} {\bf 8} (2020).

\bibitem{asp2022}
M.~E. Asp, {\it et~al.\/}, {\it PNAS Nexus\/} {\bf 1}, pgac025 (2022).

\bibitem{witmann2023}
R.~Wittmann, N.~G, A.~nd~Löwen, F.~J. Schwarzendahl, A.~Sengupta, {\it Communications Physics (in press)\/}  (2023).

\bibitem{han2021}
S.~Han, {\it et~al.\/}, {\it Soil Biology and Biochemistry\/} {\bf 154}, 108143 (2021).

\bibitem{ranjard2001quantitative}
L.~Ranjard, A.~Richaume, {\it Research in microbiology\/} {\bf 152}, 707 (2001).

\bibitem{mannik2009bacterial}
J.~M{\"a}nnik, R.~Driessen, P.~Galajda, J.~E. Keymer, C.~Dekker, {\it Proceedings of the National Academy of Sciences\/} {\bf 106}, 14861 (2009).

\bibitem{grundmann2004spatial}
G.~L. Grundmann, {\it FEMS microbiology ecology\/} {\bf 48}, 119 (2004).

\bibitem{dexter1988advances}
A.~R. Dexter, {\it Soil and tillage research\/} {\bf 11}, 199 (1988).

\bibitem{raynaud2014spatial}
X.~Raynaud, N.~Nunan, {\it PloS one\/} {\bf 9}, e87217 (2014).

\bibitem{gantner2006situ}
S.~Gantner, {\it et~al.\/}, {\it FEMS microbiology ecology\/} {\bf 56}, 188 (2006).

\bibitem{battin2016ecology}
T.~J. Battin, K.~Besemer, M.~M. Bengtsson, A.~M. Romani, A.~I. Packmann, {\it Nature Reviews Microbiology\/} {\bf 14}, 251 (2016).

\bibitem{najafi2019swimming}
J.~Najafi, F.~Altegoer, G.~Bange, C.~Wagner, {\it Soft matter\/} {\bf 15}, 10029 (2019).

\bibitem{bargul2016species}
J.~L. Bargul, {\it et~al.\/}, {\it PLoS pathogens\/} {\bf 12}, e1005448 (2016).

\bibitem{araujo2023}
N.~A. Araújo, a.~et, {\it Soft Matter\/} {\bf 19}, 1695 (2023).

\bibitem{hazas2022}
M.~B. Hazas, F.~Ziliotto, M.~Rolle, G.~Chiogna, {\it Physical Review E\/} {\bf 105}, 035105 (2022).

\bibitem{tecon2016bacterial}
R.~Tecon, D.~Or, {\it Scientific reports\/} {\bf 6}, 1 (2016).

\bibitem{or2007physical}
D.~Or, B.~F. Smets, J.~Wraith, A.~Dechesne, S.~Friedman, {\it Advances in Water Resources\/} {\bf 30}, 1505 (2007).

\bibitem{dechesne2010hydration}
A.~Dechesne, G.~Wang, G.~G{\"u}lez, D.~Or, B.~F. Smets, {\it Proceedings of the National Academy of Sciences\/} {\bf 107}, 14369 (2010).

\bibitem{rivkina2000metabolic}
E.~Rivkina, E.~Friedmann, C.~McKay, D.~Gilichinsky, {\it Applied and environmental microbiology\/} {\bf 66}, 3230 (2000).

\bibitem{wadhwa2022bacterial}
N.~Wadhwa, H.~C. Berg, {\it Nature reviews microbiology\/} {\bf 20}, 161 (2022).

\bibitem{thormann2022wrapped}
K.~M. Thormann, C.~Beta, M.~J. K{\"u}hn, {\it Annual Review of Microbiology\/} {\bf 76}, 349 (2022).

\bibitem{yonekura2003complete}
K.~Yonekura, S.~Maki-Yonekura, K.~Namba, {\it Nature\/} {\bf 424}, 643 (2003).

\bibitem{grognot2021}
M.~Grognot, K.~M. Taute, {\it Current Opinion in Microbiology\/} {\bf 61}, 73 (2021).

\bibitem{schavemaker2022flagellar}
P.~E. Schavemaker, M.~Lynch, {\it Elife\/} {\bf 11}, e77266 (2022).

\bibitem{lauga2016bacterial}
E.~Lauga, {\it Annual Review of Fluid Mechanics\/} {\bf 48}, 105 (2016).

\bibitem{figueroa20203d}
N.~Figueroa-Morales, {\it et~al.\/}, {\it Physical Review X\/} {\bf 10}, 021004 (2020).

\bibitem{cates2013active}
M.~E. Cates, J.~Tailleur, {\it Europhysics Letters\/} {\bf 101}, 20010 (2013).

\bibitem{keegstra2022ecological}
J.~M. Keegstra, F.~Carrara, R.~Stocker, {\it Nature Reviews Microbiology\/} {\bf 20}, 491 (2022).

\bibitem{mattingly2021escherichia}
H.~Mattingly, K.~Kamino, B.~Machta, T.~Emonet, {\it Nature physics\/} {\bf 17}, 1426 (2021).

\bibitem{grognot2021multiscale}
M.~Grognot, K.~M. Taute, {\it Communications biology\/} {\bf 4}, 669 (2021).

\bibitem{colin2019chemotactic}
R.~Colin, K.~Drescher, V.~Sourjik, {\it Nature communications\/} {\bf 10}, 5329 (2019).

\bibitem{fu2018spatial}
X.~Fu, {\it et~al.\/}, {\it Nature communications\/} {\bf 9}, 2177 (2018).

\bibitem{bhattacharjee2022chemotactic}
T.~Bhattacharjee, D.~B. Amchin, R.~Alert, J.~A. Ott, S.~S. Datta, {\it Elife\/} {\bf 11}, e71226 (2022).

\bibitem{adler1966chemotaxis}
J.~Adler, {\it Science\/} {\bf 153}, 708 (1966).

\bibitem{budrene1995dynamics}
E.~O. Budrene, H.~C. Berg, {\it Nature\/} {\bf 376}, 49 (1995).

\bibitem{bhattacharjee2021chemotactic}
T.~Bhattacharjee, D.~B. Amchin, J.~A. Ott, F.~Kratz, S.~S. Datta, {\it Biophysical Journal\/} {\bf 120}, 3483 (2021).

\bibitem{cremer2019chemotaxis}
J.~Cremer, {\it et~al.\/}, {\it Nature\/} {\bf 575}, 658 (2019).

\bibitem{ni2020growth}
B.~Ni, R.~Colin, H.~Link, R.~G. Endres, V.~Sourjik, {\it Proceedings of the National Academy of Sciences\/} {\bf 117}, 595 (2020).

\bibitem{neumann2021high}
R.~F. Neumann, {\it et~al.\/}, {\it Scientific reports\/} {\bf 11}, 11370 (2021).

\bibitem{kearns2010field}
D.~B. Kearns, {\it Nature Reviews Microbiology\/} {\bf 8}, 634 (2010).

\bibitem{partridge2020tumble}
J.~D. Partridge, N.~T. Nhu, Y.~S. Dufour, R.~M. Harshey, {\it Mbio\/} {\bf 11}, e01189 (2020).

\bibitem{zhang2010upper}
R.~Zhang, L.~Turner, H.~C. Berg, {\it Proceedings of the National Academy of Sciences\/} {\bf 107}, 288 (2010).

\bibitem{wang2005sensing}
Q.~Wang, A.~Suzuki, S.~Mariconda, S.~Porwollik, R.~M. Harshey, {\it The EMBO journal\/} {\bf 24}, 2034 (2005).

\bibitem{engelhardt2022novel}
I.~Engelhardt, {\it et~al.\/}, {\it The ISME Journal\/} {\bf 16}, 2337 (2022).

\bibitem{mattick2002type}
J.~S. Mattick, {\it Annual review of microbiology\/} {\bf 56}, 289 (2002).

\bibitem{craig2019type}
L.~Craig, K.~T. Forest, B.~Maier, {\it Nature reviews microbiology\/} {\bf 17}, 429 (2019).

\bibitem{merz2000pilus}
A.~J. Merz, M.~So, M.~P. Sheetz, {\it Nature\/} {\bf 407}, 98 (2000).

\bibitem{zhao2013psl}
K.~Zhao, {\it et~al.\/}, {\it Nature\/} {\bf 497}, 388 (2013).

\bibitem{oliveira2016single}
N.~M. Oliveira, K.~R. Foster, W.~M. Durham, {\it Proceedings of the National Academy of Sciences\/} {\bf 113}, 6532 (2016).

\bibitem{marathe2014bacterial}
R.~Marathe, {\it et~al.\/}, {\it Nature communications\/} {\bf 5}, 1 (2014).

\bibitem{pfreundt2023}
U.~Pfreundt, {\it et~al.\/}, {\it Science\/} {\bf 380}, 830 (2023).

\bibitem{faluweki2023}
M.~K. Faluweki, J.~Cammann, M.~G. Mazza, G.~L, {\it Phys. Rev. Lett.\/} {\bf 131}, 158303 (2023).

\bibitem{dell2018growing}
D.~Dell’Arciprete, {\it et~al.\/}, {\it Nature communications\/} {\bf 9}, 1 (2018).

\bibitem{dhar2022self}
J.~Dhar, A.~L. Thai, A.~Ghoshal, L.~Giomi, A.~Sengupta, {\it Nature Physics\/} {\bf 18}, 945 (2022).

\bibitem{rani2023}
R.~Garima, S.~Anupam, {\it bioRxiv\/} {\bf 10.1101/2023.09.07.556749} (2023).

\bibitem{figueroa2020coli}
N.~Figueroa-Morales, {\it et~al.\/}, {\it Science advances\/} {\bf 6}, eaay0155 (2020).

\bibitem{kim2015colonization}
A.~Siryaporn, M.~K. Kim, Y.~Shen, Z.~Gitai, H.~A. Stone, {\it Biophysical Journal\/} {\bf 108}, 599a (2015).

\bibitem{grau2015duo}
R.~R. Grau, {\it et~al.\/}, {\it MBio\/} {\bf 6}, e00581 (2015).

\bibitem{kinsinger2003rapid}
R.~F. Kinsinger, M.~C. Shirk, R.~Fall, {\it Journal of bacteriology\/} {\bf 185}, 5627 (2003).

\bibitem{meacock2021bacteria}
O.~J. Meacock, A.~Doostmohammadi, K.~R. Foster, J.~M. Yeomans, W.~M. Durham, {\it Nature Physics\/} {\bf 17}, 205 (2021).

\bibitem{wheeler2019not}
J.~D. Wheeler, E.~Secchi, R.~Rusconi, R.~Stocker, {\it Annual review of cell and developmental biology\/} {\bf 35}, 213 (2019).

\bibitem{zheng2021implication}
S.~Zheng, {\it et~al.\/}, {\it Frontiers in Bioengineering and Biotechnology\/} {\bf 9}, 643722 (2021).

\bibitem{liu2011idling}
J.~Liu, R.~M. Ford, J.~A. Smith, {\it Environmental science \& technology\/} {\bf 45}, 3945 (2011).

\bibitem{creppy2019effect}
A.~Creppy, E.~Cl{\'e}ment, C.~Douarche, M.~V. d'Angelo, H.~Auradou, {\it Physical Review Fluids\/} {\bf 4}, 013102 (2019).

\bibitem{rusconi2014microfluidics}
R.~Rusconi, M.~Garren, R.~Stocker, {\it Annual review of biophysics\/} {\bf 43}, 65 (2014).

\bibitem{rusconi2014bacterial}
R.~Rusconi, J.~S. Guasto, R.~Stocker, {\it Nature physics\/} {\bf 10}, 212 (2014).

\bibitem{altshuler2013flow}
E.~Altshuler, {\it et~al.\/}, {\it Soft Matter\/} {\bf 9}, 1864 (2013).

\bibitem{mino2018coli}
G.~L. Mi{\~n}o, {\it et~al.\/}, {\it Advances in Microbiology\/} {\bf 8}, 451 (2018).

\bibitem{lee2021influence}
M.~Lee, C.~Lohrmann, K.~Szuttor, H.~Auradou, C.~Holm, {\it Soft Matter\/} {\bf 17}, 893 (2021).

\bibitem{secchi2020effect}
E.~Secchi, {\it et~al.\/}, {\it Nature communications\/} {\bf 11}, 1 (2020).

\bibitem{mathijssen2019oscillatory}
A.~J. Mathijssen, {\it et~al.\/}, {\it Nature communications\/} {\bf 10}, 1 (2019).

\bibitem{marcos2012bacterial}
Marcos, H.~C. Fu, T.~R. Powers, R.~Stocker, {\it Proceedings of the National Academy of Sciences\/} {\bf 109}, 4780 (2012).

\bibitem{jing2020chirality}
G.~Jing, A.~Z{\"o}ttl, {\'E}.~Cl{\'e}ment, A.~Lindner, {\it Science advances\/} {\bf 6}, eabb2012 (2020).

\bibitem{dey2022oscillatory}
R.~Dey, C.~M. Buness, B.~V. Hokmabad, C.~Jin, C.~C. Maass, {\it Nature communications\/} {\bf 13}, 1 (2022).

\bibitem{thomas2002bacterial}
W.~E. Thomas, E.~Trintchina, M.~Forero, V.~Vogel, E.~V. Sokurenko, {\it Cell\/} {\bf 109}, 913 (2002).

\bibitem{thomas2004shear}
W.~E. Thomas, L.~M. Nilsson, M.~Forero, E.~V. Sokurenko, V.~Vogel, {\it Molecular microbiology\/} {\bf 53}, 1545 (2004).

\bibitem{thomas2008biophysics}
W.~E. Thomas, V.~Vogel, E.~Sokurenko, {\it Annu. Rev. Biophys.\/} {\bf 37}, 399 (2008).

\bibitem{zhong2017transport}
H.~Zhong, {\it et~al.\/}, {\it Biotechnology advances\/} {\bf 35}, 490 (2017).

\bibitem{weaver2012fluid}
W.~M. Weaver, V.~Milisavljevic, J.~F. Miller, D.~Di~Carlo, {\it Applied and Environmental Microbiology\/} {\bf 78}, 5890 (2012).

\bibitem{rodesney2017mechanosensing}
C.~A. Rodesney, {\it et~al.\/}, {\it Proceedings of the National Academy of Sciences\/} {\bf 114}, 5906 (2017).

\bibitem{ganesan2016associative}
M.~Ganesan, S.~Knier, J.~G. Younger, M.~J. Solomon, {\it Macromolecules\/} {\bf 49}, 8313 (2016).

\bibitem{ma2006analysis}
L.~Ma, K.~D. Jackson, R.~M. Landry, M.~R. Parsek, D.~J. Wozniak, {\it Journal of bacteriology\/} {\bf 188}, 8213 (2006).

\bibitem{drescher2013biofilm}
K.~Drescher, Y.~Shen, B.~L. Bassler, H.~A. Stone, {\it Proceedings of the National Academy of Sciences\/} {\bf 110}, 4345 (2013).

\bibitem{friedlander2013bacterial}
R.~S. Friedlander, {\it et~al.\/}, {\it Proceedings of the National Academy of Sciences\/} {\bf 110}, 5624 (2013).

\bibitem{duffy1995random}
K.~J. Duffy, P.~T. Cummings, R.~M. Ford, {\it Biophysical journal\/} {\bf 68}, 800 (1995).

\bibitem{zeitz2017active}
M.~Zeitz, K.~Wolff, H.~Stark, {\it The European Physical Journal E\/} {\bf 40}, 1 (2017).

\bibitem{reichhardt2014active}
C.~Reichhardt, C.~O. Reichhardt, {\it Physical Review E\/} {\bf 90}, 012701 (2014).

\bibitem{chepizhko2013diffusion}
O.~Chepizhko, F.~Peruani, {\it Physical review letters\/} {\bf 111}, 160604 (2013).

\bibitem{morin2017diffusion}
A.~Morin, D.~L. Cardozo, V.~Chikkadi, D.~Bartolo, {\it Physical Review E\/} {\bf 96}, 042611 (2017).

\bibitem{spagnolie2023swimming}
S.~E. Spagnolie, P.~T. Underhill, {\it Annual Review of Condensed Matter Physics\/} {\bf 14}, 381 (2023).

\bibitem{volpe2017topography}
G.~Volpe, G.~Volpe, {\it Proceedings of the National Academy of Sciences\/} {\bf 114}, 11350 (2017).

\bibitem{licata2016diffusion}
N.~A. Licata, B.~Mohari, C.~Fuqua, S.~Setayeshgar, {\it Biophysical journal\/} {\bf 110}, 247 (2016).

\bibitem{bertrand2018optimized}
T.~Bertrand, Y.~Zhao, O.~B{\'e}nichou, J.~Tailleur, R.~Voituriez, {\it Physical Review Letters\/} {\bf 120}, 198103 (2018).

\bibitem{irani2022dynamics}
E.~Irani, Z.~Mokhtari, A.~Zippelius, {\it Physical Review Letters\/} {\bf 128}, 144501 (2022).

\bibitem{wolfe1989migration}
A.~J. Wolfe, H.~C. Berg, {\it Proceedings of the National Academy of Sciences\/} {\bf 86}, 6973 (1989).

\bibitem{bhattacharjee2019bacterial}
T.~Bhattacharjee, S.~S. Datta, {\it Nature communications\/} {\bf 10}, 1 (2019).

\bibitem{bhattacharjee2019confinement}
T.~Bhattacharjee, S.~S. Datta, {\it Soft matter\/} {\bf 15}, 9920 (2019).

\bibitem{perez2021impact}
L.~J. Perez, T.~Bhattacharjee, S.~S. Datta, R.~Parashar, N.~L. Sund, {\it Physical Review E\/} {\bf 103}, 012611 (2021).

\bibitem{kurzthaler2021geometric}
C.~Kurzthaler, {\it et~al.\/}, {\it Nature communications\/} {\bf 12}, 1 (2021).

\bibitem{dehkharghani2023self}
A.~Dehkharghani, N.~Waisbord, J.~S. Guasto, {\it Communications Physics\/} {\bf 6}, 18 (2023).

\bibitem{chang2018surface}
Y.-R. Chang, E.~R. Weeks, W.~A. Ducker, {\it ACS applied materials \& interfaces\/} {\bf 10}, 9225 (2018).

\bibitem{brown2016swimming}
A.~T. Brown, {\it et~al.\/}, {\it Soft matter\/} {\bf 12}, 131 (2016).

\bibitem{makarchuk2019enhanced}
S.~Makarchuk, V.~C. Braz, N.~A. Ara{\'u}jo, L.~Ciric, G.~Volpe, {\it Nature Communications\/} {\bf 10}, 4110 (2019).

\bibitem{o2016sensational}
G.~A. O’Toole, G.~C. Wong, {\it Current opinion in microbiology\/} {\bf 30}, 139 (2016).

\bibitem{yang2022bacterial}
K.~Yang, {\it et~al.\/}, {\it Journal of Materials Science \& Technology\/} {\bf 99}, 82 (2022).

\bibitem{geng2014bacteria}
J.~Geng, C.~Beloin, J.-M. Ghigo, N.~Henry, {\it PLoS One\/} {\bf 9}, e102049 (2014).

\bibitem{chu2018self}
E.~K. Chu, O.~Kilic, H.~Cho, A.~Groisman, A.~Levchenko, {\it Nature communications\/} {\bf 9}, 4087 (2018).

\bibitem{lynch2022transitioning}
J.~B. Lynch, {\it et~al.\/}, {\it Biophysical Journal\/} {\bf 121}, 2653 (2022).

\bibitem{yin2022escaping}
Y.~Yin, {\it et~al.\/}, {\it Biophysical Journal\/} {\bf 121}, 4656 (2022).

\bibitem{carnes2010confinement}
E.~C. Carnes, {\it et~al.\/}, {\it Nature chemical biology\/} {\bf 6}, 41 (2010).

\bibitem{kuhn2017bacteria}
M.~J. K{\"u}hn, F.~K. Schmidt, B.~Eckhardt, K.~M. Thormann, {\it Proceedings of the National Academy of Sciences\/} {\bf 114}, 6340 (2017).

\bibitem{borer2021}
B.~Borer, D.~Or, {\it Current Opinion in Biotechnology\/} {\bf 67}, 65 (2021).

\bibitem{jansson2023}
J.~K. Jansson, R.~McClure, R.~G. Egbert, {\it Nature Biotechnology\/} pp. 1--13 (2023).

\bibitem{cichos2020}
F.~Cichos, K.~Gustavsson, B.~Mehlig, G.~Volpe, {\it Nat Mach Intell.\/} {\bf 2}, 94–103 (2020).

\bibitem{meacock2023}
O.~J. Meacock, W.~M. Durham, {\it PLOS Computational Biology\/} {\bf 19}, e1011524 (2023).

\bibitem{vamsee2023}
V.~Ulaganathan, A.~Sengupta, {\it Journal of Colloid and Interface Science\/} {\bf 649}, 302 (2023).

\bibitem{sengupta2013}
A.~Sengupta, {\it International Journal of Molecular Sciences\/} {\bf 14}, 22826 (2013).

\end{thebibliography}

\bibliographystyle{Science}

\newpage



\end{document}